\documentclass[10pt,onecolumn,aps,prd,preprintnumbers,showpacs,superscriptaddress,nofootinbib,amsmath,amssymb,floats,floatfix,showkeys,notitlepage,longbibliography]{revtex4-2}
\usepackage{lineno}

\usepackage{orcidlink}
\usepackage{comment}
\usepackage{lipsum}
\usepackage{graphicx}
\usepackage{subfigure}
\usepackage{palatino}
\usepackage{sans}
\usepackage{hyperref}
\hypersetup{colorlinks=true,linkcolor=blue,urlcolor=blue,citecolor=blue}
\usepackage[toc,page]{appendix}
\usepackage[normalem]{ulem}
\usepackage{adjustbox}
\usepackage{latexsym}
\usepackage{amsmath}
\numberwithin{equation}{section}
\usepackage{amssymb}
\usepackage{amsfonts}
\usepackage{dcolumn}
\usepackage{bm}
\usepackage{tikz}
\usetikzlibrary{decorations.pathmorphing}
\usepackage{bigints}
\usepackage{array,tabularx,multirow,booktabs}
\usepackage[tracking=true]{microtype}
\usepackage{soul} 
\SetTracking{}{500}
\SetTracking{encoding={*}, shape=sc}{40}
\allowdisplaybreaks

\usepackage{xcolor} 

\renewcommand{\Re}{\ensuremath{\mathrm{Re}}}

\begin{document} \sloppy

\title{Time-dependent black hole lensing from ringdown quasinormal mode}

\author{Reggie C. Pantig \orcidlink{0000-0002-3101-8591}} 
\email{rcpantig@mapua.edu.ph}
\affiliation{Physics Department, School of Foundational Studies and Education, Map\'ua University, 658 Muralla St., Intramuros, Manila 1002, Philippines.}

\begin{abstract}
Is it possible to find imprints of a black hole ringdown through gravitational lensing? To address this question, we formulate an analytic description of weak-field and strong-deflection lensing of light in a time-dependent, perturbed Schwarzschild spacetime. The spacetime dynamics are modeled by a single, axisymmetric, even-parity quasinormal mode with \(\ell=2\), \(m=0\) and complex frequency \(\omega\). Working to first order in a small perturbation amplitude while keeping background null geodesics exact, we derive a time-dependent line-of-sight (Born) expression for the screen-plane deflection measured by a static observer at large radius. From the same integral, an asymptotic expansion yields the familiar weak-field \(1/b\) law with a ringdown-frequency correction that drives a harmonic centroid wobble, whereas a near-photon-sphere expansion produces a time-dependent generalization of the logarithmic strong-deflection limit with modulated coefficients, including a small oscillation of the critical impact parameter. An observer tetrad built from the background static frame ensures that all screen-plane quantities, such as centroid motion, multi-image hierarchy, and time delays, as well as photon-ring morphology, are gauge-safe at first order. We provide explicit matching across regimes, showing that the near-critical coefficients governing spacing and ring-radius modulations are encoded in the same Born kernel that controls the weak-field correction. This provides an analytic account of how ringdown-scale perturbations enter imaging observables, without resorting to numerical integration of null geodesics.
\end{abstract}

\pacs{04.70.Bw; 04.25.Nx; 04.30.-w; 98.62.Sb; 95.30.Sf; 04.20.-q}
\keywords{Gravitational lensing; black-hole ringdown; quasinormal modes; Zerilli perturbations; strong-deflection limit; centroid wobble}

\maketitle

\section{Introduction} \label{sec1}

Gravitational lensing by compact objects remains a powerful probe of strong-field gravity and the nature of dark compact sources \cite{Zhong:2024ysg,Johnson:2019ljv,Kaiser:1996wk,Damour:1998jm,Kopeikin:1999ev,Tsukamoto:2016jzh,Vagnozzi:2022moj,Allahyari:2019jqz,Khodadi:2020jij,Molla:2025yoh,Ditta:2025vsa,Turakhonov:2024xfg,Alloqulov:2024sns,Wang:2025fmz}. Foundational analyses of Schwarzschild lensing and its extensions to naked singularities established the basic phenomenology of multiple images, relativistic loops, and their observational signatures \cite{Virbhadra:1999nm,Virbhadra:2002ju}. Early work also highlighted how additional fields can imprint themselves on lensing observables \cite{Virbhadra:1998dy} and connected these effects to broader questions such as Seifert's conjecture \cite{Virbhadra:1998kd}. The differential‐geometric underpinning of photon surfaces furnished a rigorous framework for locating unstable light rings that control both strong lensing and shadow formation \cite{Claudel:2000yi}. Building on this, detailed predictions for relativistic images, time delays, and magnification centroids were developed and refined \cite{Virbhadra:2008ws,Virbhadra:2007kw}. Recent progress has quantified image distortions in Schwarzschild lensing and uncovered conservation properties of distortion measures, providing robust diagnostics that are less sensitive to astrophysical systematics \cite{Virbhadra:2022iiy,Virbhadra:2024xpk}. In parallel, precise relations between the photon-sphere radius and black-hole shadow have been extended to include cosmological constant corrections, clarifying how large-scale curvature perturbs near-horizon optics \cite{Adler:2022qtb}. These theoretical advances feed directly into constraints on the compactness of supermassive dark objects at galactic centers, sharpening tests of black-hole paradigms with high-resolution imaging and timing data \cite{Virbhadra:2022ybp}. Realistic black hole lensing is rarely vacuum: ambient plasma and magnetic fields induce frequency-dependent deflection, reshape caustics, and shift photon rings and shadows \cite{Morozova:2013uyv,Schee:2017hof,Turimov:2018ttf}. Such chromatic and dispersive effects persist across rotating and regular spacetimes, complicating image interpretation \cite{Morozova:2013uyv,Schee:2017hof}. Beyond GR, braneworld geometries predict systematic deviations in strong-deflection and retrolensing observables, offering complementary tests \cite{Abdujabbarov:2017pfw}. More broadly, parametrized and quantum-gravity-motivated black-hole geometries continue to motivate joint studies of shadow observables and quasinormal spectra, further underscoring the value of analytic tools that connect optical and ringdown diagnostics \cite{Konoplya:2021slg,Bonanno:2025dry}.

Gravitational lensing by black holes spans a wide dynamic range from weak deflections in the asymptotic region to the strong-field regime near the photon sphere \cite{Virbhadra:1999nm,Perlick:2021aok}. In parallel, the spacetime itself can be time-dependent, for instance, during black-hole ringdown when quasinormal modes (QNMs) drive metric perturbations with characteristic complex frequencies \cite{Berti:2009kk,Kokkotas:1999bd,Konoplya:2011qq}. These two features, wide spatial scales and intrinsic temporal modulation, are often modeled with separate tools that obscure their continuity. Our objective is to provide a single, purely analytical framework that treats both regimes within one description and that projects directly onto the observer's screen in a gauge-safe manner.

We concentrate on light propagation in the Schwarzschild geometry of mass \(M\), perturbed at first order by a single axisymmetric even-parity Zerilli mode with angular numbers \(\ell=2\), \(m=0\) and complex frequency \(\omega=\omega_R+i\omega_I\) with \(\omega_I<0\) \cite{Zerilli:1970wzz,Moncrief:1974am}. For the lowest Schwarzschild overtones, the ringdown scale is set by \(M^{-1}\) (see standard QNM tables in Ref. \cite{Berti:2009kk}). This choice provides the simplest dynamical driver of ringdown that preserves axial symmetry, minimizes technical overhead, and still captures the essential physics of a decaying spacetime oscillation. Throughout, we work perturbatively in a small, dimensionless amplitude \(\varepsilon\) and keep the unperturbed null geodesics exact, so that all time dependence arises from the interaction of the QNM with the photon's background trajectory \cite{Kopeikin:1999ev,Lewis:2006fu}.

Our central construct is a time-dependent line-of-sight map that gives the deflection measured by a static observer at a large areal radius. The map is evaluated along the unperturbed null geodesic labeled by its impact parameter \(b\). In the far field, it reproduces the familiar weak-bending law ($1/b$) \cite{Weinberg:1972kfs,Will:2014kxa} with an \(\varepsilon\)-suppressed, frequency-resolved correction that produces a harmonic motion of the image centroid on the screen. In the near-critical regime \(b\to b_c^+\), with \(b_c\) the Schwarzschild critical impact parameter, the same map yields a time-dependent generalization of the logarithmic strong-deflection relation \cite{Bozza:2002zj,Tsukamoto:2016jzh}. The coefficients of this relation become slowly varying functions of the observer's time, all oscillating at the QNM frequency. The critical scale itself acquires a small modulation that controls the most singular response near the photon ring \cite{Chandrasekhar:1985kt,Johnson:2019ljv,Gralla:2020srx,EventHorizonTelescope:2019dse,EventHorizonTelescope:2022wkp}.

A key structural feature of our approach is the use of an observer tetrad fixed by the background static frame \cite{Moncrief:1974am,Johnson:2019ljv}. Screen angles are constructed directly from locally measured photon momenta. At first order in \(\varepsilon\), this guarantees that centroid shifts, image positions, and ring morphology are insensitive to gauge transformations that would otherwise act on the metric perturbation. In this way, the imaging diagnostics we develop are bona fide observables at the order of interest.

Recent work has begun to probe time-dependent gravitational lensing during black-hole relaxation. In particular, Zhong, Cardoso, and Chen \cite{Zhong:2024ysg} have used backward ray tracing in a relaxing Kerr background to show that the photon deflection inherits the characteristic ringdown pattern at intermediate times and exhibits a \(t_o^{-3}\) late-time decay, thereby establishing the qualitative imprint of QNM dynamics on light propagation \cite{Price:1971fb}. Earlier analyses of time-dependent deflection by gravitational waves or localized, retarded sources also demonstrated oscillatory signatures in lensing and timing, albeit in regimes far from the photon sphere and without a black-hole ringdown setting \cite{Kaiser:1996wk,Damour:1998jm,Wang:2019skw,Pantig:2024kfn}. At the same time, the strong-deflection limit (SDL) for stationary spacetimes is well understood, including the logarithmic divergence and its coefficients (\(\bar a,\bar b,b_c\)) that control image hierarchies and photon-ring morphology \cite{Bozza:2002zj,Tsukamoto:2016jzh}. Classic studies of light propagation in nonstationary fields also delineate key limits: in particular, for localized sources, the wave-zone and intermediate-zone contributions cancel so that the leading time-dependent deflection is near-zone and falls as \(b^{-3}\) (contrasting with the \(1/b\) Born imprint we obtain from a global QNM on a black-hole background), while stochastic or plane-wave backgrounds and moving lenses generate distinctive, velocity-dependent deflection and delay terms in the weak field \cite{Kaiser:1996wk,Damour:1998jm,Kopeikin:1999ev}. Finally, oscillatory near-horizon structures can drive ring-scale astrometric modulations akin to ours, for example, superradiant bosonic clouds around spinning holes, which likewise predict QNM-frequency screen signatures in photon-ring observables, though sourced by matter fields rather than by the spacetime's own even-parity QNM \cite{Chen:2022kzv}.

A central goal of this paper is to express screen-plane lensing observables for a time-dependent perturbation in a form that applies from the weak-field regime to the strong-deflection limit. We do so by deriving a single line-of-sight Born kernel that is valid for all impact parameters at \(\mathcal{O}(\varepsilon)\), and by extracting both the far-field \(1/b\) correction and a time-dependent extension of the SDL through controlled asymptotic expansions. The same kernel also governs the modulation of the critical scale \(b_c(t_o)\) and the associated near-critical imaging diagnostics. For the gravitational-wave analogue, namely, the image-type-dependent distortions of lensed GW waveforms even in the geometric-optics limit, see Ref. \cite{Dai:2017huk}. Such a treatment allows us to track how a single QNM perturbation induces a shared phase in weak-field centroid motion and in near-critical photon-ring modulations, with both regimes governed by analytic asymptotics of the same line-of-sight kernel.

The presentation is organized as follows. Section \ref{sec2} specifies the background Schwarzschild geometry, the \(\ell=2,m=0\) even-parity perturbation through the Zerilli master field, and the observer tetrad together with the screen construction and gauge considerations. Section \ref{sec3} derives the time-dependent line-of-sight map and develops its far-field expansion, which produces the centroid wobble at the QNM frequency. Section \ref{sec4} extracts the near-photon-sphere behavior and obtains the time-dependent strong-deflection law together with the hierarchy of relativistic images and their arrival-time structure. Section \ref{sec5} performs the matching between the weak and strong limits and assembles the screen-plane observables into a coherent set of diagnostics that are continuous across regimes. Section \ref{sec6} summarizes the framework and outlines natural extensions. Technical material is collected in three appendices: Appendix \ref{appA} reviews the Zerilli formalism and reconstructs the metric perturbation; Appendix \ref{appB} derives the master Born kernel and its equivalent representations; Appendix \ref{appC} develops the near-critical expansion and expresses the strong-deflection coefficients in terms of localized kernel data at the photon sphere.

\section{Background, Perturbation, and Observers} \label{sec2}
In this section, we set up the spacetime, perturbation content, and observational frame used throughout. Our background is the Schwarzschild geometry of mass \(M\) in geometrized units \(G=c=1\). We then introduce a first-order, axisymmetric, even-parity QNM perturbation with angular numbers \((\ell,m)=(2,0)\) and complex frequency \(\omega=\omega_R+i\omega_I\) with \(\omega_I<0\). The perturbation is described by the Zerilli master function, whose dynamics and boundary conditions fix the time dependence that subsequently enters our lensing maps. The observer frame and screen construction will be given in Sec. \ref{subsec2.2}.

\subsection{Schwarzschild + even-parity \texorpdfstring{\((\ell,m)=(2,0)\)}{} QNM} \label{subsec2.1}
We take as background the Schwarzschild line element \cite{Schwarzschild:1916uq,Wald:1984rg,Chandrasekhar:1985kt}
\begin{equation}
    ds^2 = -f(r)dt^2 + f(r)^{-1}dr^2 + r^2\left(d\theta^2+\sin^2\theta d\phi^2\right), \label{1}
\end{equation}
where 
\begin{equation}
    f(r)=1-\frac{2M}{r},
\end{equation}
Here, \(t\) is the Schwarzschild coordinate time, \(r\) is the areal coordinate radius, and \((\theta,\phi)\) are the standard spherical coordinate angles. A null geodesic will be parameterized by an affine parameter \(\lambda\) with four-momentum \(p_\mu=dx_\mu/d\lambda\), though in this subsection we focus on the spacetime rather than the geodesic equations.

We consider a linear perturbation of amplitude \(\varepsilon\ll1\) \cite{Regge:1957td,Martel:2005ir},
\begin{equation}
    g_{\mu\nu}(t,r,\theta,\phi) = g^{(0)}_{\mu\nu}(r) + \varepsilon\,h_{\mu\nu}(t,r,\theta,\phi), \label{2}
\end{equation}
where \(g^{(0)}_{\mu\nu}\) is the Schwarzschild metric in Eq. \eqref{1} and \(h_{\mu\nu}\) is an even-parity, axisymmetric QNM with \((\ell,m)=(2,0)\). In the Regge-Wheeler-Zerilli (RWZ) gauge for \(\ell\ge2\), the nonvanishing components of an even-parity, axisymmetric perturbation can be written as \cite{Regge:1957td,Zerilli:1970wzz,Moncrief:1974am,Martel:2005ir}
\begin{align}
h_{tt} &= f H_0(t,r)\,P_2(\cos\theta), \nonumber \\
h_{tr} &= H_1(t,r)\,P_2(\cos\theta), \nonumber \\
h_{rr} &= f^{-1}H_2(t,r)\,P_2(\cos\theta), \nonumber \\
h_{\theta\theta} &= r^2 K(t,r)\,P_2(\cos\theta), \nonumber \\
h_{\phi\phi} &= r^2 K(t,r)\,P_2(\cos\theta)\,\sin^2\theta, \label{3}
\end{align} 
where \(H_0,H_1,H_2,K\) are functions of \((t,r)\) and \(P_2\) is the Legendre polynomial of degree \(\ell=2\). Axisymmetry implies no \(\phi\)-dependence. Throughout the axisymmetric ($m=0$) analysis here, we use the Legendre basis $P_\ell(\cos\theta)$ for angular dependence. In particular, for $\ell=2$, $P_2(0)=-1/2$ for an equatorial observer. It is also convenient to package \((H_0,H_1,H_2,K)\) into the gauge-invariant Zerilli master function \(\Psi_Z(t,r)\), which encodes the dynamical content of the even-parity sector at first order.

Introducing the tortoise coordinate \(r_*\) defined by \cite{Chandrasekhar:1985kt}
\begin{equation}
\frac{dr_*}{dr} = f(r)^{-1}, \qquad r_* = r + 2M \ln\,\left(\frac{r}{2M}-1\right), \label{4}
\end{equation}
the Zerilli function obeys the wave equation
\begin{equation}
\left(-\partial_t^2 + \partial_{r_*}^2 - V_Z(r)\right)\Psi_Z(t,r) = 0, \label{5}
\end{equation}
with the Zerilli potential \cite{Zerilli:1970wzz,Moncrief:1974am}
\begin{align}
V_Z(r) &= \frac{2 f(r)}{r^3}\,
\frac{\lambda^2(\lambda+1)r^3 + 3\lambda^2 M r^2 + 9\lambda M^2 r + 9 M^3}{\left(\lambda r + 3M\right)^2},
\nonumber \\
\lambda &\equiv \frac{(\ell-1)(\ell+2)}{2}. \label{6}
\end{align}
For the \((\ell,m)=(2,0)\) QNM, we impose homogeneous boundary conditions consistent with black-hole ringdown: purely ingoing at the event horizon and purely outgoing at null infinity. Writing \( \Psi_Z(t,r) \propto e^{-i\omega t} \psi(r) \) with complex frequency \(\omega=\omega_R+i\omega_I\) and \(\omega_I<0\), the asymptotics are \cite{Kokkotas:1999bd,Berti:2009kk,Vishveshwara:1970cc,Leaver:1985ax}
\begin{align}
\Psi_Z &\sim e^{-i\omega(t+r_*)} \quad \text{for} \quad (r\to 2M), \nonumber \\
\Psi_Z &\sim e^{-i\omega(t-r_*)} \quad \text{for} \quad (r\to \infty). \label{7}
\end{align}
These conditions quantize \(\omega\) to the discrete \(\ell=2\) even-parity spectrum and fix the temporal decay \(e^{\omega_I t}\) inherent to ringdown.

At linear order in \(\varepsilon\), the metric perturbation \(h_{\mu\nu}\) can be reconstructed from \(\Psi_Z\) through algebraic and first-derivative relations standard in the RWZ formalism; we defer the explicit formulas to Appendix \ref{appA}. For definiteness, we normalize \(\Psi_Z\) so that the leading outgoing amplitude at large \(r\) is unity, and we absorb the physical smallness of the perturbation into \(\varepsilon\). With these conventions, Eq. \eqref{2} provides a definite axisymmetric, even-parity \(\ell=2\) driving mode that sources the time dependence used in our lensing construction in later sections.

\subsection{Observer tetrad and gauge} \label{subsec2.2}
We model detection at future null infinity by an asymptotic static observer located at areal radius \(r_o\gg M\) and polar angle \(\theta_o\). The observer's four-velocity \(u^\mu\) is aligned with the static Killing field at zeroth order and is corrected at \(\mathcal{O}(\varepsilon)\) so that normalization holds in the perturbed geometry of Eq. \eqref{2}. A convenient orthonormal tetrad \(e^{\mu}{}_{\hat{a}}\) \((\hat{a}=\hat{0},\hat{r},\hat{\theta},\hat{\phi})\) at the observer worldline is \cite{Wald:1984rg,Chandrasekhar:1985kt,Frittelli:1999yf}
\begin{align}
e^{\mu}{}_{\hat{0}} &= f(r_o)^{-1/2}\,\delta^\mu{}_t, \nonumber \\
e^{\mu}{}_{\hat{r}} &= f(r_o)^{1/2}\,\delta^\mu{}_r, \nonumber \\ 
e^{\mu}{}_{\hat{\theta}} &= \frac{1}{r_o}\,\delta^\mu{}_\theta, \nonumber \\
e^{\mu}{}_{\hat{\phi}} &= \frac{1}{r_o\sin\theta_o}\,\delta^\mu{}_\phi, \label{8}
\end{align}
with \(f\) given in Eq. \eqref{1} (here $\theta_o$ is the polar angle of the observer; the $\hat{\phi}$-leg contains $\sin\theta_o$ as usual). These vectors satisfy \cite{Cunningham_1972,Perlick:2021aok}
\begin{equation}
g_{\mu\nu}(x_o)\,e^{\mu}{}_{\hat{a}}\,e^{\nu}{}_{\hat{b}}=\eta_{\hat{a}\hat{b}} + \mathcal{O}(\varepsilon^2), \label{9}
\end{equation}
where \(x_o^\mu\) denotes the observer event and \(\eta_{\hat{a}\hat{b}}=\mathrm{diag}(-1,1,1,1)\). The \(\mathcal{O}(\varepsilon)\) corrections implicit in Eq. \eqref{9} are uniquely fixed by enforcing orthonormality with the full metric of Eq. \eqref{2}; their explicit forms are deferred to Appendix \ref{appA}.

\begin{figure}
    \centering
    \includegraphics[width=0.60\textwidth]{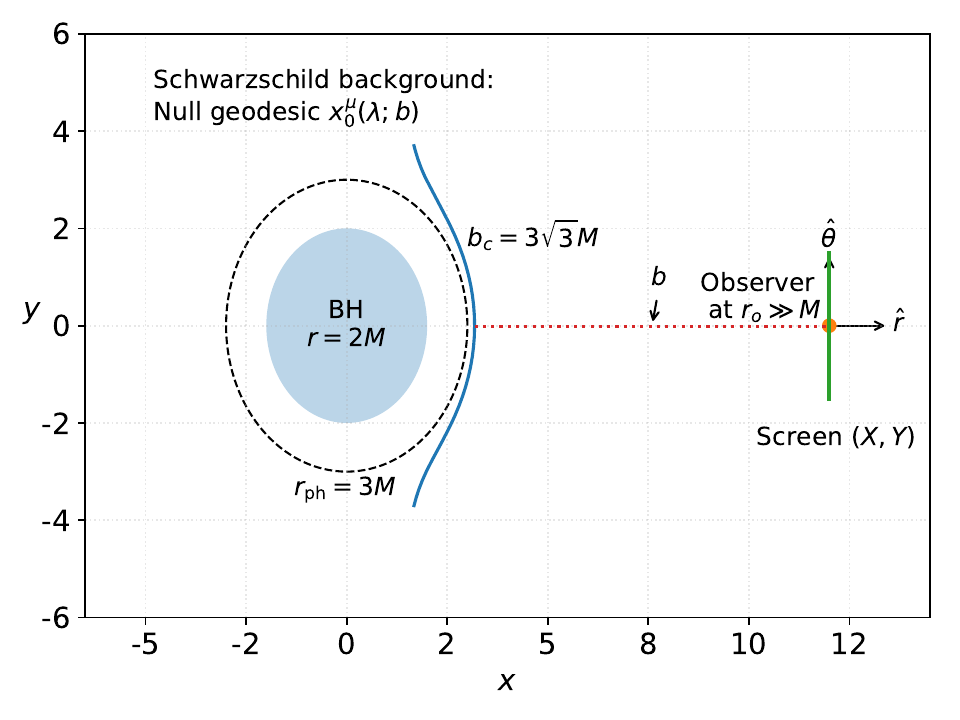}
    \caption{Hybrid geometry-plus-screen schematic used to fix conventions. The Schwarzschild spatial geometry is represented by the horizon at $r=2M$ (filled disk), the photon sphere at $r_{\rm ph}=3M$ (dashed circle), and a representative near-critical null trajectory with closest approach $r_0(b)$. The asymptotic observer at $r_o\gg M$ carries the tetrad introduced in Sec. \ref{subsec2.2} and measures the local screen coordinates $(X,Y)$ on the observer plane. The quantity $b_c=3\sqrt{3}\,M$ is the Schwarzschild \emph{critical impact parameter} associated with the near-critical ray family; it is not a separate curve on the diagram. The figure is intended only as a convention-fixing schematic rather than as a literal embedding of the observer screen into the spacetime picture. This setup underlies the background bending integral in Eq. \eqref{16} and the time-dependent Born map in Eqs. \eqref{18}-\eqref{22}.}
    \label{fig1}
\end{figure}

Let \(p^\mu\) be the photon four-momentum at \(x_o^\mu\). The components measured in the observer frame are
\begin{equation}
p^{\hat{a}} = e^{\hat{a}}{}_{\mu}\,p^\mu, \qquad e^{\hat{a}}{}_{\mu}\,e^{\mu}{}_{\hat{b}}=\delta^{\hat{a}}{}_{\hat{b}}, \label{10}
\end{equation}
and the instantaneous celestial direction is encoded in the two-vector
\begin{equation}
\boldsymbol{n} \equiv \frac{\boldsymbol{p}}{p^{\hat{0}}}
= \left(\frac{p^{\hat{r}}}{p^{\hat{0}}},\frac{p^{\hat{\theta}}}{p^{\hat{0}}},\frac{p^{\hat{\phi}}}{p^{\hat{0}}}\right), \label{11}
\end{equation}
where bold symbols denote spatial components in the tetrad. We define the observer screen as the plane orthogonal to \(e^{\mu}{}_{\hat{r}}\) at \(x_o^\mu\) and adopt the right-handed screen basis \((\hat{X},\hat{Y})=(\hat{\phi},-\hat{\theta})\). The screen coordinates \((X,Y)\) of the photon are obtained by the usual perspective map for a local inertial observer \cite{Cunningham_1972,Schneider_1992,Perlick:2021aok,Johnson:2019ljv},
\begin{equation}
X = -\frac{r_o\,p^{\hat{\phi}}}{p^{\hat{0}}}, \qquad
Y = \frac{r_o\,p^{\hat{\theta}}}{p^{\hat{0}}}, \qquad
b \equiv \sqrt{X^2+Y^2}, \label{12}
\end{equation}
which defines the impact parameter \(b\) measured on the screen. Eq. \eqref{12} is equivalent to the common celestial-sphere parametrization by small angles \((\theta_X,\theta_Y)=(X/r_o,Y/r_o)\) at large \(r_o\). We adopt the convention that $+X$ points toward decreasing $\phi$ (hence the minus sign in $X$), and $+Y$ toward increasing $\theta$. For clarity, Fig. \ref{fig1} summarizes both the background ray geometry and the local observer-screen conventions in a single schematic.
Figure \ref{fig1} is a hybrid schematic: it combines the Schwarzschild spatial geometry with the observer's local screen construction in a single convention-fixing diagram. It should therefore not be read as a literal image-plane plot or as an embedding of the screen into the spacetime geometry. Its purpose is simply to display, in one place, the horizon, the photon sphere, a representative near-critical ray, the observer location, and the orientation of the screen coordinates \((X,Y)\). In particular, the label \(b_c=3\sqrt{3}\,M\) denotes the critical impact parameter of the near-critical null family, while the photon sphere at \(r_{\rm ph}=3M\) highlights the domain \(b\to b_c^+\) central to Secs. \ref{sec4}-\ref{sec5}. All weak- and strong-field observables discussed later are measured on the observer's screen defined in Eq. \eqref{12}.

Because the perturbation is axisymmetric, we may, without loss of generality, take \(\theta_o=\pi/2\) for most calculations. At zeroth order in \(\varepsilon\) the conserved energy \(E=-p_t\) and angular momentum \(L=p_\phi\) of the photon in the Schwarzschild background Eq. \eqref{1} satisfy
\begin{equation}
\frac{X}{r_o} \simeq -\frac{L}{E\,r_o\sin\theta_o}, \quad
\frac{Y}{r_o} \simeq \frac{p_\theta}{E\,r_o}, \quad
b = \frac{L}{E} + \mathcal{O}\,\left(\frac{M}{r_o}\right), \label{13}
\end{equation}
so that \(b\) reduces to the usual ratio \(L/E\) in the asymptotically flat region. In later sections, we shall describe deflection through a two-dimensional vector \(\boldsymbol{\alpha}(b,t_o)\) defined on the screen; its weak- and strong-field limits follow from the line-of-sight Born representation derived in Sec. \ref{sec3}.

We now address gauge issues at first order in \(\varepsilon\). A linearized diffeomorphism generated by a vector field \(\xi^\mu=\mathcal{O}(\varepsilon)\) acts on the metric perturbation as
\begin{equation}
h_{\mu\nu}\rightarrow h_{\mu\nu} - \nabla_\mu \xi_\nu - \nabla_\nu \xi_\mu, \label{14}
\end{equation}
and on tensorial quantities derived from \(h_{\mu\nu}\) accordingly. We implement two safeguards to ensure that screen observables built from Eq. \eqref{12} are gauge-independent at \(\mathcal{O}(\varepsilon)\). First, all dynamical content of the even-parity perturbation is encoded in the Zerilli gauge-invariant master field \(\Psi_Z\) of Eqs. \eqref{5}-\eqref{7}, and the reconstruction used in Appendix \ref{appA} expresses \(p^{\hat{a}}\) corrections in terms of \(\Psi_Z\) and its derivatives only. Second, the tetrad is fixed operationally by the normalization and orthogonality conditions in Eq. \eqref{9} at the physical observer event \(x_o^\mu\). With these choices, any change induced by Eq. \eqref{14} in the local frame is a pure \(\mathcal{O}(\varepsilon)\) Lorentz transformation that leaves \((X,Y)\) invariant at that order. Concretely, one finds that the gauge variation of the screen map is a boundary term proportional to \(\xi^\mu\) evaluated at \(r_o\), and our asymptotic conditions impose \(\xi^\mu=\mathcal{O}(1/r_o)\), hence
\begin{equation}
\delta_{\xi}X=\delta_{\xi}Y=0+\mathcal{O}\,\left(\frac{\varepsilon}{r_o}\right). \label{15}
\end{equation}
Eq. \eqref{15} guarantees that the centroid motion, multi-image separations, and photon-ring modulations defined later are bona fide observables at first order.

Finally, we denote by \(t_o\) the Schwarzschild time coordinate of the detection event \(x_o^\mu\), and by \(\lambda_o\) the corresponding affine parameter value on the photon worldline. The association \(t\mapsto t_o\) defines the phase with which the QNM signal in Eqs. \eqref{5}-\eqref{7} is sampled by the observer and will be the sole time variable entering the deflection \(\boldsymbol{\alpha}(b,t_o)\) in Secs. \ref{sec3}-\ref{sec5}.

\section{Unified Born Integral and Weak-Field Expansion} \label{sec3}
We now derive a single, time-dependent line-of-sight expression for the lensing deflection that is valid to first order in the perturbation amplitude \(\varepsilon\) of Eq. \eqref{2} while keeping the Schwarzschild background of Eq. \eqref{1} exact. The resulting master Born integral will be evaluated along the unperturbed null geodesic with impact parameter \(b\) and will furnish, by asymptotic expansion, both the far-field Born law (Sec. \ref{subsec3.2}) and the time-dependent strong-deflection form (Sec. \ref{sec4}). All observable angles are defined on the screen introduced in Sec. \ref{subsec2.2} and are therefore gauge-safe to \(\mathcal{O}(\varepsilon)\).

\subsection{Master time-dependent Born integral} \label{subsec3.1}
Let \(x_0^\mu(\lambda,b)\) be the null geodesic of the Schwarzschild background (Eq. \eqref{1}) with conserved energy \(E=-p_{0t}\) and angular momentum \(L=p_{0\phi}\), and let \(p_0^\mu(\lambda)=dx_0^\mu/d\lambda\) be its four-momentum. The impact parameter measured at infinity is \(b=L/E\) to leading order, consistent with Eq. \eqref{13}. The background bending angle \(\alpha_0(b)\) is given exactly by the standard integral \cite{Weinberg:1972kfs}
\begin{equation}
\alpha_0(b)=2\int_{r_0(b)}^{\infty}\frac{dr}{r^2}\,
\left[\frac{1}{b^2}-\frac{f(r)}{r^2}\right]^{-1/2} - \pi, \label{16}
\end{equation}
where \(r_0(b)\) is the distance of closest approach determined by \(b^2=r_0^2/f(r_0)\).

We switch on the first-order, axisymmetric, even-parity QNM perturbation of Eq. \eqref{2}, encoded by the Zerilli master field \(\Psi_Z\) that satisfies Eqs. \eqref{5}-\eqref{7}. In the Born or single-scattering approximation on the curved background, the photon follows \(x_0^\mu(\lambda,b)\) while its momentum acquires an \(\mathcal{O}(\varepsilon)\) correction from the perturbed Hamiltonian \(H=(1/2) g^{\mu\nu}p_\mu p_\nu=0\). Linearizing Hamilton's equation \(dp_\mu/d\lambda=-\partial_\mu H\) about the background solution and evaluating the source on \(x_0^\mu\) and \(p_{0}^\mu\) yields
\begin{equation}
\delta p_{\mu}(\lambda_o,b) = -\frac{1}{2}\int_{\lambda_s}^{\lambda_o}\, d\lambda\,
\partial_\mu h_{\alpha\beta}\left(x_0(\lambda)\right)\,p_0^\alpha(\lambda)\,p_0^\beta(\lambda),
\label{17}
\end{equation}
where \(\lambda_o\) and \(\lambda_s\) denote, respectively, the values at the observer and the source, and indices are raised/lowered with the background metric. Eq. \eqref{17} is the covariant Born response for the photon momentum on the fixed Schwarzschild trajectory.

The approximation in Eq. \eqref{17} treats the perturbation to first order in \(\varepsilon\) while evaluating it along the unperturbed Schwarzschild null ray. It is controlled by requiring that the induced change in the momentum remain perturbative along the integration domain, i.e., that \(|\delta p_\mu|/|p_{0\mu}|\ll 1\) for the rays of interest. In the near-critical regime \(b\to b_c^{+}\), the prolonged dwell time close to the photon sphere enhances the accumulated response and therefore tightens the effective bound on \(\varepsilon\); this refinement is made explicit in Sec. \ref{sec4} when we take the near-photon-sphere limit of the same kernel.

Projecting \(\delta p_\mu\) on the observer tetrad of Eq. \eqref{8} and converting to screen angles via Eq. \eqref{12} gives the master deflection map. At the observer event \(x_o^\mu\), the perturbed momentum is first converted to tetrad components,
\begin{equation}
p^{\hat a}=e^{\hat a}{}_{\mu}p^\mu
= p_0^{\hat a}+\varepsilon\,\delta p^{\hat a},
\qquad
\delta p^{\hat a}=e^{\hat a}{}_{\mu}\,\delta p^\mu ,
\label{eq:main-tetrad-variation}
\end{equation}
where the \(\mathcal{O}(\varepsilon)\) correction to the tetrad contributes only at \(\mathcal{O}(\varepsilon^2)\) when combined with \(\delta p^\mu\). Varying the screen map Eq. \eqref{12} at fixed \(x_o^\mu\) then yields
\begin{equation}
\delta\alpha_A
=
\frac{\delta p^{\widehat A}}{p_0^{\hat0}}
-\frac{p_0^{\widehat A}}{(p_0^{\hat0})^2}\,\delta p^{\hat0}
=
\frac{1}{p_0^{\hat0}}\,\Pi_A{}^{\ \hat b}\,\delta p_{\hat b},
\qquad A\in\{X,Y\},
\label{eq:main-screen-projector}
\end{equation}
where \(\Pi_A{}^{\ \hat b}\) is the projector onto the observer screen,
\begin{equation}
\Pi_A{}^{\ \hat b}
=
\delta_A{}^{\ \hat b}
-\alpha_{0,A}\,\delta^{\hat b}{}_{\hat0},
\qquad
\alpha_{0,A}\equiv \frac{p_0^{\widehat A}}{p_0^{\hat0}}.
\label{eq:main-screen-projector-def}
\end{equation}
Substituting Eq. \eqref{17} into Eq. \eqref{eq:main-screen-projector}, and using
\(\delta p_{\hat b}=e_{\hat b}{}^{\ \mu}\delta p_\mu\), gives the master deflection map
\begin{equation}
\boldsymbol{\alpha}(b,t_o)=\alpha_0(b) + \varepsilon\,\delta\boldsymbol{\alpha}(b,t_o), \label{18}
\end{equation}
with \(\alpha_0(b)\) determined by Eq. \eqref{16}, and
\begin{equation}
\delta\alpha_A(b, t_o)=\int_{\lambda_s}^{\lambda_o}\, d\lambda\,
\mathcal{K}_A\,\left[x_0(\lambda),p_0(\lambda)\right],\qquad A\in X,Y . \label{19}
\end{equation}
The kernel \(\mathcal{K}_A\) is therefore nothing more than the screen-projected linear response of the photon momentum, evaluated along the unperturbed null geodesic. A convenient representative is
\begin{align}
\mathcal{K}_A\,\left[x_0(\lambda),p_0(\lambda)\right]
&=-\frac{1}{2\,p_0^{\hat{0}}(\lambda_o)}\,
\Pi_{A}{}^{\ \hat{b}}\,e_{\hat{b}}{}^{\ \mu}(x_o)\, \nonumber \\
&\times \partial_\mu h_{\alpha\beta}\left(x_0(\lambda)\right)\,
p_0^\alpha(\lambda)\,p_0^\beta(\lambda).
\label{20}
\end{align}
The projector form in Eq. \eqref{20} is one convenient representative; Appendix \ref{appB} shows its equivalence to a manifestly covariant \(\delta\Gamma\) line integral with cancelling boundary terms under our asymptotic conditions. Different but equivalent kernel representatives follow from integrating Eq. \eqref{17} by parts and using the background geodesic equation; all yield the same \(\delta\boldsymbol{\alpha}\) at \(\mathcal{O}(\varepsilon)\). Equations \eqref{eq:main-tetrad-variation}--\eqref{20} make explicit that the image-plane deflection is obtained by varying the screen map itself, rather than being introduced as an additional assumption.

For later reference, the derivation chain is
\[
H \;\longrightarrow\; \delta p_\mu \;\longrightarrow\; \delta p^{\hat a}
\;\longrightarrow\; \delta\alpha_A \;\longrightarrow\; \mathcal{K}_A .
\]
Appendix \ref{appB} provides the same sequence in fully explicit form, including the equivalent \(\delta\Gamma\) representation and the cancellation of endpoint terms. The QNM time dependence in Eqs. \eqref{5}-\eqref{7} implies that, along the background ray, the perturbation takes the form
\begin{align}
h_{\mu\nu}\left(x_0(\lambda)\right)
&=\mathrm{Re}\left\{\hat{h}_{\mu\nu}\left(r(\lambda),\theta(\lambda)\right)\,e^{-i\omega\,t_0(\lambda)}\right\}, \nonumber \\
\omega&=\omega_R+i\omega_I \quad \text{for} \quad \omega_I<0, \label{21}
\end{align}
where \(t_0(\lambda)\) is the Schwarzschild time along \(x_0^\mu(\lambda)\). Writing \(t_0(\lambda)=t_o-\tau(\lambda)\) with \(\tau(\lambda)\ge0\) the light-travel delay between the encounter point and the observer event, Eq. \eqref{19} can be cast as
\begin{align}
&\delta\alpha_A(b,t_o)=
\mathrm{Re}\left\{e^{-i\omega t_o}\,
\mathcal{A}_A(b,\omega)\right\}, \nonumber \\
&\mathcal{A}_A(b,\omega)=\int_{\lambda_s}^{\lambda_o}\, d\lambda\,
\tilde{\mathcal{K}}_A\,\left[r(\lambda),\theta(\lambda)\right]\,
e^{i\omega\,\tau(\lambda)}, \label{22}
\end{align}
where \(\tilde{\mathcal{K}}_A\) is the complex kernel obtained from \(\mathcal{K}_A\) after factoring out the harmonic time dependence (the explicit form follows from Appendix \ref{appB} together with the RWZ reconstruction in Appendix \ref{appA}). Eq. \eqref{22} shows that the measured deflection inherits the single QNM frequency \(\omega\); its amplitude \(\mathcal{A}_A\) depends on \(b\) only through the unperturbed trajectory and the delay profile \(\tau(\lambda)\).

In Eq. \eqref{22}, we interpret the QNM factor \(e^{-i\omega t}\) as the ringdown component of a causal perturbation excited at finite retarded time, rather than as a globally defined normal mode on all of spacetime. Accordingly, the line-of-sight integral is to be understood under the same asymptotic assumptions used to discard boundary terms (Appendix \ref{appB}): effectively, the integrand has compact support or sufficiently rapid falloff along the ray, so that no additional endpoint contributions arise beyond \(\mathcal{O}(\varepsilon/r_o)\).

Eqs. \eqref{18}-\eqref{22} constitute the master, time-dependent Born formalism. Two regimes will be obtained from the same expression. In Sec. \ref{subsec3.2} we expand Eq. \eqref{22} for \(M/b\ll 1\) to recover the far-field Born law and the centroid wobble. In Sec. \ref{sec4} we analyze the contribution from segments of the trajectory that probe \(r\approx 3M\), showing that Eq. \eqref{22} reproduces the logarithmic strong-deflection behavior once \(b\to b_c^{+}=3\sqrt{3}M\), with the time dependence inherited from the same phase factor.

\subsection{Far-field (Born) expansion and centroid wobble} \label{subsec3.2}
We now evaluate Eqs. \eqref{18}-\eqref{22} in the weak-deflection domain \(M/b\ll1\), keeping the Schwarzschild background exact along the ray but expanding the integrals in powers of \(M/b\). The background bending reduces to \cite{Weinberg:1972kfs,Will:2014kxa,Schneider_1992}
\begin{equation}
\alpha_0(b)=\frac{4M}{b}+\mathcal{O}\,\left(\frac{M^2}{b^2}\right), \label{23}
\end{equation}
as obtained from Eq. \eqref{16}. For the perturbation, we use Eq. \eqref{22} with the unperturbed trajectory approximated by a straight line of impact parameter \(b\) through the asymptotically flat region, so that
\begin{equation}
r(\lambda)\simeq \sqrt{b^2+z^2},\qquad t_0(\lambda)=t_o-\tau(\lambda),\qquad \tau(\lambda)\simeq z, \label{24}
\end{equation}
where \(z\) is a signed longitudinal coordinate along the asymptotic straight ray, chosen so that \(z=0\) at the point of closest approach; the source and observer correspond to \(z\to-\infty\) and \(z\to+\infty\), respectively. In this approximation the delay \(\tau(\lambda)\) differs from \(z\) by an additive constant fixed by \(\tau(\lambda_o)=0\) and by a branch-dependent sign; these conventions only affect the overall complex normalization and can be absorbed into \(c_A(\omega)\) in Eq. \eqref{25}.

In the far zone the even-parity \(\ell=2\) QNM behaves as an outgoing quadrupolar wave with \(h_{\mu\nu}\sim r^{-1}e^{-i\omega(t-r_*)}\), hence \(\partial_\mu h_{\alpha\beta}\sim r^{-2}\,e^{-i\omega(t-r_*)}\). Substituting this falloff into Eq. \eqref{19} or, equivalently, Eq. \eqref{22}, we obtain the leading Born correction
\begin{equation}
\delta\alpha_A(b,t_o)
= \mathrm{Re}\left\{e^{-i\omega t_o}\,\frac{c_A(\omega)}{b}\,\mathcal{F}_A(\omega b)\right\}
+\mathcal{O}\,\left(\frac{M}{b^2}\right), \qquad A\in{X,Y}, \label{25}
\end{equation}
where \(c_A(\omega)\) are complex, frequency-dependent coefficients fixed by the RWZ reconstruction (Appendix \ref{appA}) and the kernel choice (Appendix \ref{appB}), and
\begin{equation}
\mathcal{F}_A(\omega b)\equiv \int_{-\infty}^{+\infty}\,
\frac{dz}{b}\,\widehat{\mathcal{K}}_A\,\left(\frac{\sqrt{b^2+z^2}}{b}\right)\,e^{i\omega z}  \label{26}
\end{equation}
is a dimensionless Fresnel-type factor encoding the finite-frequency sampling of the QNM along the nearly straight path. The explicit form of \(\widehat{\mathcal{K}}_A\) is given in Appendix \ref{appB}. Two limits of Eq. \eqref{25} are especially transparent:
\begin{equation}
\delta\alpha_A(b,t_o)\sim
\begin{cases}
\displaystyle \mathrm{Re}\left\{e^{-i\omega t_o}\,\frac{\tilde c_A(\omega)}{b}\right\}, & |\omega|\,b\ll 1,\\
\displaystyle \mathrm{Re}\left\{e^{-i\omega t_o}\,\frac{\hat c_A(\omega)}{b}\,e^{-|\mathrm{Im}\omega|\,b}\right\}, & |\omega|\,b\gg 1,
\end{cases}
\label{27}
\end{equation}
with \(\tilde c_A,\hat c_A\) calculable constants. Thus the leading time-dependent correction decays as \(1/b\) and inherits the QNM phase \(e^{-i\omega t_o}\), in agreement with the structure of Eq. \eqref{22}.

Collecting Eqs. \eqref{23} and \eqref{25}, the far-field deflection vector on the screen reads
\begin{align}
\boldsymbol{\alpha}(b,t_o)&=\frac{4M}{b}\,\hat{\boldsymbol{e}}_b
+\varepsilon\,\mathrm{Re}\left\{e^{-i\omega t_o}\,\frac{\boldsymbol{\mathcal{A}}(b,\omega)}{b}\right\} \nonumber \\
&+\mathcal{O}\,\left(\frac{M^2}{b^2},\varepsilon\frac{M}{b^2}\right), \label{28}
\end{align}
where \(\hat{\boldsymbol{e}}_b\) is the outward radial unit vector in the \((X,Y)\) plane and \(\boldsymbol{\mathcal{A}}=(c_X\mathcal{F}_X,c_Y\mathcal{F}_Y)\). That is, on the observer's screen we use the orthonormal basis $\{\hat{\boldsymbol{e}}_{X},\hat{\boldsymbol{e}}_{Y}\}$ and define the screen radial unit vector $\hat{\boldsymbol{e}}_{b}\equiv (X\,\hat{\boldsymbol{e}}_{X}+Y\,\hat{\boldsymbol{e}}_{Y})/b$. The small \(\mathcal{O}(M/b^2)\) terms arise from curvature corrections to Eq. \eqref{24} and from subleading pieces of the kernel in Eq. \eqref{19}.
\begin{figure}
    \centering
    \includegraphics[width=0.60\textwidth]{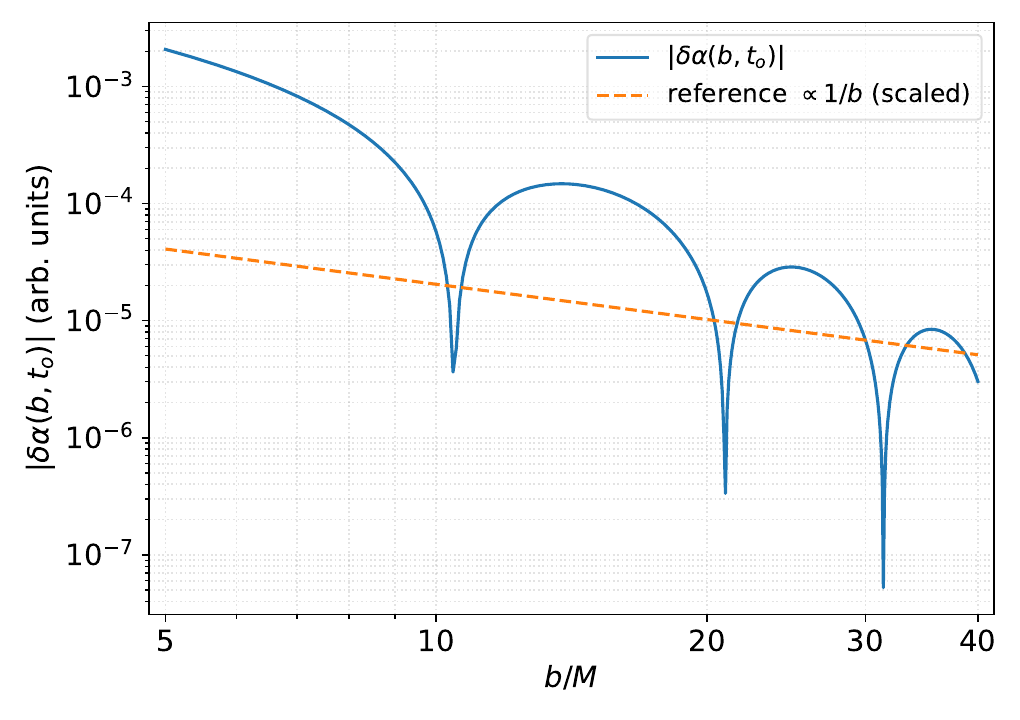}
    \caption{Far-field, QNM-driven deflection amplitude vs impact parameter. The $\mathcal{O}(\varepsilon)$ Born correction obeys $|\delta\alpha(b,t_o)|\sim |A(b,\omega)|/b$ with mild oscillations from the Fresnel factor $F_A(\omega b)$ [Eqs. \eqref{25}-\eqref{28}]. Using logarithmic axes makes the $1/b$ decay explicit (slope $-1$). A dashed curve shows a reference $\propto 1/b$ law scaled at a pivot $b_{\rm ref}$ for comparison.}
    \label{fig8}
\end{figure}

As seen in Fig. \ref{fig8}, the $\mathcal{O}(\varepsilon)$ deflection decays like $1/b$, with gentle structure governed by $\mathcal{F}_A(\omega b)$. This is the weak-field imprint that, when averaged over a bundle, yields the centroid wobble in Eq. \eqref{30}.

To connect with observables we project Eq. \eqref{28} onto the screen angles \((\theta_X,\theta_Y)=(X/r_o,Y/r_o)\) defined in Eq. \eqref{12}. For a narrow bundle of rays of fixed \(b\) arriving at \((X,Y)\), the apparent position shift is
\begin{equation}
\Delta\boldsymbol{\theta}(b,t_o)=\frac{\boldsymbol{\alpha}(b,t_o)}{r_o}, \label{29}
\end{equation}
up to the usual thin-deflection identification valid at \(r_o\gg M\). Equivalently, introducing the complex image-plane coordinate
\begin{equation}
\zeta(b,t_o)\equiv \frac{X+iY}{r_o}
=\frac{\alpha_X(b,t_o)}{r_o}+i\,\frac{\alpha_Y(b,t_o)}{r_o},
\label{29a}
\end{equation}
shows that the line-of-sight Born map provides a direct map from the ray label \(b\) and observer time \(t_o\) to a point on the observer's screen. In this sense, the formalism already reconstructs image-plane kinematics, rather than merely identifying asymptotic structure in the deflection.

We define the image centroid \(\boldsymbol{\theta}_c(t_o)\) as the intensity-weighted average position of the bundle on the screen; within the present purely theoretical treatment, we capture its dynamics by the average of Eq. \eqref{29} over the bundle's narrow impact-parameter distribution. The centroid, therefore, exhibits a harmonic wobble at the QNM frequency:
\begin{equation}
\boldsymbol{\theta}_c(t_o)=\boldsymbol{\theta}_{c,0}
+\varepsilon\,\mathrm{Re}\left\{e^{-i\omega t_o}\,\frac{\boldsymbol{\mathcal{B}}(\omega)}{r_o}\right\}
+\mathcal{O}\,\left(\frac{M}{r_o}\frac{M}{b^2}\right), \label{30}
\end{equation}
where \(\boldsymbol{\theta}_{c,0}\) is the static centroid from \(\alpha_0\) and \(\boldsymbol{\mathcal{B}}(\omega)\) is a bundle-dependent complex amplitude obtained from \(\boldsymbol{\mathcal{A}}(b,\omega)\) in Eq. \eqref{28}. The phase of the wobble equals \(\arg \boldsymbol{\mathcal{B}}\) and is inherited from the travel-time weighting in Eq. \eqref{22}. The only additional input here is the bundle weight used in the centroid average; the screen position of each individual ray is already fixed by Eqs. \eqref{29}-\eqref{29a}.
\begin{figure}
    \centering
    \includegraphics[width=0.60\textwidth]{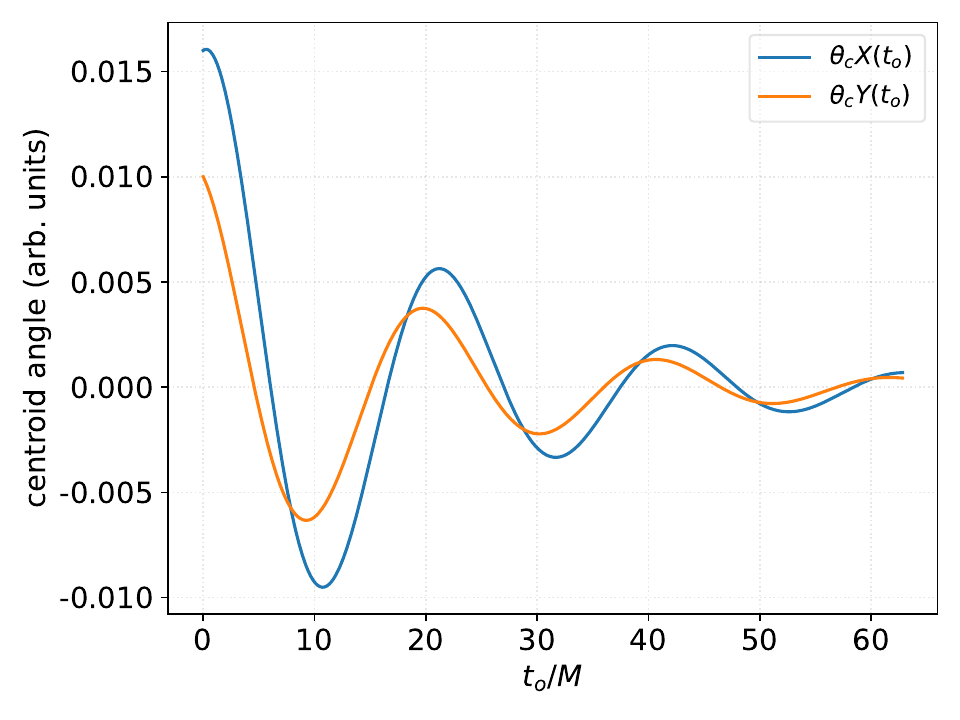}
    \caption{Centroid wobble in the weak field from Eq. \eqref{30}. The components $\theta_{c,X}(t_o)$ and $\theta_{c,Y}(t_o)$ execute a damped harmonic motion at the QNM frequency $\omega=\omega_R+i\omega_I$ with $\omega_I<0$, with complex amplitude set by $\mathcal{B}(\omega)$. This panel visualizes the $\mathcal{O}(\varepsilon)$, gauge-safe centroid modulation predicted by the Born expansion in Eqs. \eqref{25}-\eqref{28}.}
    \label{fig2}
\end{figure}
As seen in Fig.~\ref{fig2}, the centroid motion is a clean, single-frequency probe of the perturbation, with decay governed by $\omega_I<0$. The phase is inherited from the line-of-sight kernel in Eq. \eqref{22}, anticipating the phase-locked relations with strong-field observables established in Sec. \ref{sec5}. Eq. \eqref{30} provides the weak-field imaging diagnostic used later to match continuously to the strong-deflection modulations derived from the same master integral in Sec. \ref{sec4}.

\section{Time-Dependent Strong-Deflection Lensing} \label{sec4}
We now extract from the master Born representation in Eqs. \eqref{18}-\eqref{22} the strong-deflection limit \(SDL\) relevant to trajectories that skim the photon sphere. The analysis is controlled by the small parameter \(\Delta\equiv b/b_c-1\) with \(0<\Delta\ll1\), where \(b\) is the impact parameter defined by Eq. \eqref{12} and \(b_c\) denotes the critical value associated with the unstable circular null orbit. The time dependence enters exclusively through the \((\ell,m)=(2,0)\) QNM perturbation of Sec. \ref{sec2}, which modulates the SDL coefficients and, in particular, the critical scale \(b_c\). Our goal in this section is to obtain a time-dependent generalization of the logarithmic SDL law from the same kernel that produced the Born expansion in Sec. \ref{sec3}.

\subsection{Near-photon-sphere expansion} \label{subsec4.1}
On the Schwarzschild background, the circular photon orbit sits at the photon-sphere radius \cite{Claudel:2000yi,Cunha:2018acu}
\begin{equation}
r_{\mathrm{ph}}=3M,\qquad b_c=\frac{r_{\mathrm{ph}}}{\sqrt{f(r_{\mathrm{ph}})}}=3\sqrt{3}M. \label{31}
\end{equation}
For \(b\to b_c^{+}\) the bending angle \(\alpha_0(b)\) obtained from Eq. \eqref{16} exhibits the well-known logarithmic divergence, \cite{Bozza:2002zj,Tsukamoto:2016jzh}
\begin{equation}
\alpha_0(b)=-\bar a\,\ln\,\left(\frac{b}{b_c}-1\right)+\bar b+\mathcal{O}\,\left((b-b_c)\ln|b-b_c|\right), \label{32}
\end{equation}
with the Schwarzschild values $\bar a = 1$, and \(\bar b=\ln\,\left[216(7-4\sqrt{3})\right]-\pi\) if one adopts the standard normalization. The constant $\bar b$ depends on the subtraction convention for the $-\pi$ in the strong-deflection expansion; we follow the standard normalization commonly used for Schwarzschild. Eq. \eqref{32} follows by expanding the integral in Eq. \eqref{16} around the double root of the radial potential at \(r_{\mathrm{ph}}\).

We now switch on the perturbation of Eq. \eqref{2} and use the time-dependent Born representation, Eqs. \eqref{18}-\eqref{22}, to determine how the coefficients in Eq. \eqref{32} are modulated. Our claim is not that a generic nonstationary spacetime must obey the standard strong-deflection-limit (SDL) law. Rather, the statement here is perturbative and more specific: (i) the Schwarzschild background null geodesics are kept exact, (ii) the metric perturbation is treated only to first order in the QNM amplitude \(\varepsilon\), and (iii) the perturbation is sampled on the unperturbed rays through the line-of-sight Born kernel. Under these assumptions, the logarithmic divergence is inherited from the background photon-sphere instability, while the time dependence enters only through modulations of the corresponding coefficients and through a small shift of the critical scale.

Physically, the integral in Eq. \eqref{22} becomes dominated by the segment of the unperturbed trajectory that executes a long whirl near \(r_{\mathrm{ph}}\), where small changes in the local geometry feed into large changes of the total deflection. It is therefore convenient to introduce the near-photon-sphere coordinate
\begin{equation}
\rho \equiv \frac{r-r_{\mathrm{ph}}}{r_{\mathrm{ph}}}, \qquad |\rho|\ll 1, \label{33}
\end{equation}
and to parameterize the background null ray by the angle \(\varphi\) accumulated during the whirl phase. In this regime the unperturbed relations implied by Eq. \eqref{16} reduce to the standard form
\begin{equation}
\varphi(\rho)\simeq -\frac{1}{\sqrt{\kappa}}\,\ln|\rho|+\varphi_0, \qquad
\kappa=\frac{1}{2}r_{\mathrm{ph}}^2\,\left(\partial_r^2 V_{\mathrm{eff}}\right)_{r_{\mathrm{ph}}}, \label{34}
\end{equation}
where \(V_{\mathrm{eff}}\) is the Schwarzschild null effective potential (\(V_{\rm eff}=f\,L^2/r^2\) for equatorial motion), and \(\kappa>0\) encodes the instability exponent of the circular orbit. The logarithm in Eq. \eqref{34} is the origin of the SDL behavior in Eq. \eqref{32}: as \(b\to b_c^+\), the whirl duration scales as \(\Delta\varphi\sim -\ln\Delta\), where
\begin{equation}
\Delta \equiv \frac{b}{b_c}-1. \label{34a}
\end{equation}

To first order in the perturbation amplitude \(\varepsilon\), the metric perturbation \(h_{\mu\nu}\) sampled along the whirl contributes to the kernel in Eq. \eqref{22} through \(\partial_\mu h_{\alpha\beta} p_0^\alpha p_0^\beta\) evaluated at \(r\simeq r_{\mathrm{ph}}\). Exploiting Eq. \eqref{34} to trade the affine parameter for \(\varphi\), the correction \(\delta\boldsymbol{\alpha}\) in Eq. \eqref{18} can be written, up to terms that remain finite as \(b\to b_c^{+}\), as
\begin{align}
\delta\alpha_A(b,t_o)&=\mathrm{Re}\left\{e^{-i\omega t_o}\,
\left[\mathcal{C}_A(\omega)\,\ln\Delta +\mathcal{D}_A(\omega)\right]\right\} \nonumber \\
&+\mathcal{O}\,\left((b-b_c)\ln|b-b_c|\right), \label{35}
\end{align}
where the complex amplitudes \(\mathcal{C}_A(\omega)\) and \(\mathcal{D}_A(\omega)\) are determined by the near-photon-sphere part of the Born kernel and depend only on background quantities at \(r_{\mathrm{ph}}\) and on the QNM frequency \(\omega\). Equation \eqref{35} makes explicit that the time-dependent perturbation does \emph{not} alter the existence of the logarithmic divergence at \(\mathcal{O}(\varepsilon)\); instead, it modulates its coefficient and its finite part. The explicit expressions, obtained by evaluating the RWZ reconstruction (Appendix \ref{appA}) inside the line-of-sight integral (Appendix \ref{appB}), are not needed for the present section.

Combining Eqs. \eqref{32} and \eqref{35} with Eq. \eqref{18}, we arrive at a compact time-dependent SDL law of the form
\begin{align}
\alpha(b,t_o)&= -\bar a(t_o)\,\ln\,\left(\frac{b}{b_c(t_o)}-1\right)+\bar b(t_o) \nonumber \\
&+\mathcal{O}\,\left((b-b_c)\ln|b-b_c|\right), \label{36}
\end{align}
where
\begin{align}
\bar a(t_o)&=\bar a+\varepsilon\,\mathrm{Re}\left\{e^{-i\omega t_o}\,a_1(\omega)\right\},\nonumber \\
\bar b(t_o)&=\bar b+\varepsilon\,\mathrm{Re}\left\{e^{-i\omega t_o}\,b_1(\omega)\right\},\nonumber \\
b_c(t_o)&=b_c\left[1+\varepsilon\,\mathrm{Re}\left\{e^{-i\omega t_o}\,\beta_1(\omega)\right\}\right], \label{37}
\end{align}
with \(a_1,b_1,\beta_1\) complex response coefficients that follow from the same near-photon-sphere kernel that produced \(\mathcal{C}_A,\mathcal{D}_A\) in Eq. \eqref{35}. Eq. \eqref{36} is nothing but Eq. \eqref{32} with time-dependent coefficients. The modulation of \(b_c\) in Eq. \eqref{37} can be interpreted geometrically as a small shift of the photon-sphere location,
\begin{equation}
r_{\mathrm{ph}}(t_o)=3M+\varepsilon\,\mathrm{Re}\left\{e^{-i\omega t_o}\,\delta r_{\mathrm{ph}}(\omega)\right\}, \label{38}
\end{equation}
propagated to \(b_c(t_o)\) through the algebraic relation in Eq. \eqref{31}. The quantity \(\delta r_{\mathrm{ph}}\) is determined by the \(\mathcal{O}(\varepsilon)\) change of the null effective potential extremum in the perturbed geometry and is therefore a gauge-invariant statement at the order of interest, given our use of the Zerilli master field.

It is often convenient to linearize Eq. \eqref{36} explicitly in \(\varepsilon\). Writing \(\Delta=b/b_c-1\) and keeping only \(\mathcal{O}(\varepsilon)\) terms, we obtain
\begin{align}
\alpha(b,t_o)&=\alpha_0(b)
+\varepsilon\,\mathrm{Re}\biggl\{e^{-i\omega t_o}\biggl[-a_1(\omega)\,\ln\Delta \nonumber \\
&+\bar a\frac{\beta_1(\omega)}{\Delta}+b_1(\omega)\biggr]\biggr\}
+\mathcal{O}\,\left(\varepsilon\,\Delta\ln\Delta\right)+\mathcal{O}\,\left(\varepsilon^2\right), \label{39}
\end{align}
where \(\alpha_0(b)\) is given by Eq. \eqref{32}, and $\bar a = 1$ for the Schwarzschild case. By axisymmetry, the leading $O(\varepsilon)$ near-critical deflection is radial on the screen; we therefore drop tangential components in what follows. The term proportional to \(\beta_1/\Delta\) originates from expanding \(\ln\,\left(b/b_c(t_o)-1\right)\) and makes explicit that an oscillatory shift of \(b_c\) produces the largest time-dependent contribution as \(b\to b_c^{+}\). With our $P_2$ normalization, $\beta_1(\omega)=\frac{1}{4}\,H_0(r_{\mathrm{ph}},\omega)$. Equation \eqref{39} displays the structure of the time-dependent SDL in its most transparent linearized form. The term proportional to \(\beta_1/\Delta\) originates from expanding the shifted logarithm \(\ln\!\bigl(b/b_c(t_o)-1\bigr)\) around the static critical value \(b_c\), and therefore reflects the oscillatory displacement of the critical impact parameter itself. By contrast, the term proportional to \(-a_1\ln\Delta\) represents a genuine modulation of the logarithmic coefficient, while \(b_1\) collects the finite near-critical remainder. In this way, Eq. \eqref{39} makes explicit that the background photon-sphere instability is responsible for the divergence, whereas the QNM perturbation enters only through coefficient modulations and the shift of the critical scale.

Our first-order, line-of-sight (Born) expansion keeps the Schwarzschild background
deflection $\alpha_0(b)$ exact and expands in the QNM amplitude $\varepsilon$.
In the near-critical regime $\Delta \equiv b/b_c-1 \ll 1$, Eq. \eqref{39} shows that the
dominant time-dependent correction scales as $\delta\alpha \sim
\varepsilon\,\beta_1/\Delta$, whereas the background scales as
$\alpha_0 \sim -\ln\Delta$ [cf. Eq. \eqref{32}]. A conservative perturbative criterion
is therefore
\begin{equation}
\mathcal{R}(b)\;\equiv\;\frac{|\varepsilon\,\beta_1|}{\Delta\,|\ln\Delta|}\ll 1
\quad\Rightarrow\quad
\Delta \gg \frac{|\varepsilon\,\beta_1|}{|\ln\Delta|}\,,
\label{eq:validity-criterion}
\end{equation}
which, up to the slowly varying logarithm, is the order-of-magnitude condition
$\Delta \gtrsim \mathcal{O}(\varepsilon|\beta_1|)$. Equivalently, using the static
SDL spacing $s_n^{(0)}=\exp[(\bar b-2\pi n)/\bar a]$ [Eq. \eqref{40} with $\bar a=1$],
the $n$-th relativistic image remains within the linear regime provided
\begin{equation}
s_n^{(0)} \gg |\varepsilon\,\beta_1|
\quad\Longleftrightarrow\quad
n \lesssim \frac{1}{2\pi}\!\left[\ln\!\frac{1}{\varepsilon|\beta_1|}+\mathcal{O}(1)\right].
\end{equation}
In words: the instantaneous ring-radius wobble
$\delta\theta_{\rm ring}/\theta_c=\varepsilon\,|\beta_1|$ [Eq. \eqref{47}]
must be smaller than the image's offset from the ring
$\theta_n-\theta_{\rm ring}\simeq \theta_c\,s_n^{(0)}$.

\begin{figure}
    \centering
    \includegraphics[width=0.60\textwidth]{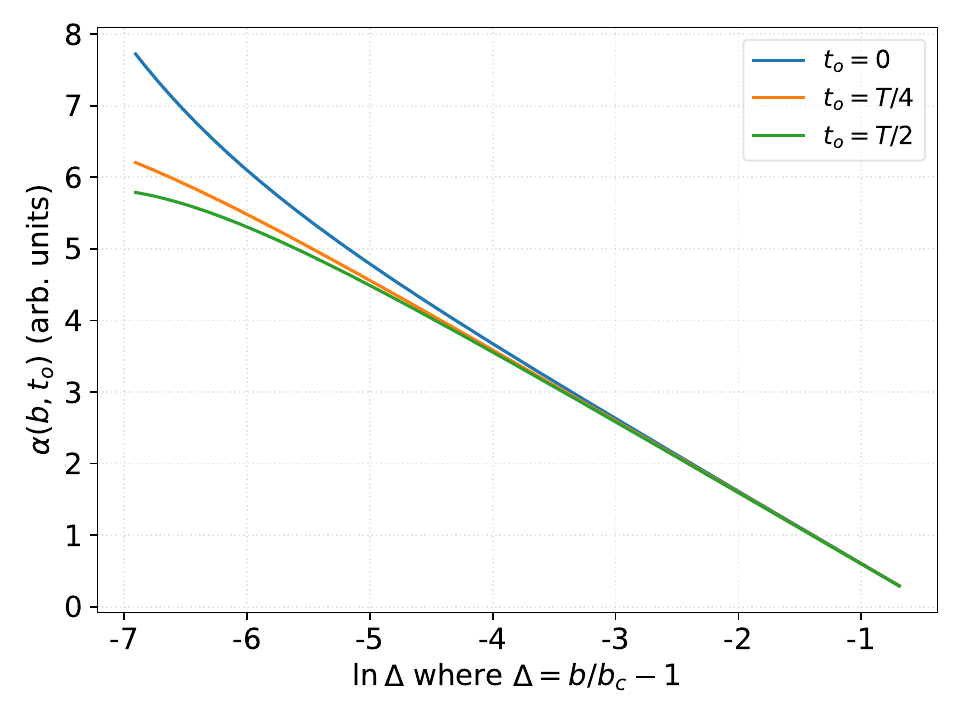}
    \caption{Near-critical deflection from the time-dependent SDL law. Shown is $\alpha(b,t_o)$ versus $\ln\Delta$ with $\Delta=b/b_c-1\ll1$, as predicted by Eq. \eqref{36} and its linearized form Eq. \eqref{39}. The dominant $1/\Delta$ sensitivity arises from the modulation of $b_c(t_o)$ through $\beta_1(\omega)$, while the slope receives a small time-dependent correction from $a_1(\omega)$. Parallel, slightly offset curves at different $t_o$ illustrate the coherent QNM-driven modulation.}
    \label{fig3}
\end{figure}

Figure \ref{fig3} makes explicit the near-critical amplification: the $\beta_1/\Delta$ piece dominates as $\Delta\to0^+$, while the $-a_1\ln\Delta$ term induces a mild slope variation. This behavior feeds directly into the image hierarchy and photon-ring modulations derived in Sec. \ref{subsec4.2}, and it will be matched to the weak-field kernel in Sec. \ref{subsec5.1}. Eq. \eqref{39} is the central near-critical result that will feed, in Sec. \ref{subsec4.2}, the logarithmic time delay, the multiplicity of relativistic images, and their photon-ring modulations, all sharing the common QNM frequency inherited from Eq. \eqref{22}.

\subsection{Time delays, multiplicity, and photon-ring modulation} \label{subsec4.2}
Eq. \eqref{36} implies that, for each winding number \(n\in\mathbb{N}\), relativistic images arise from solutions of the lens equation with total bending \cite{Bozza:2002zj,Virbhadra:2002ju,Virbhadra:2008ws} \(\alpha(b,t_o)\approx 2\pi n+\Delta\alpha_n\) where \(|\Delta\alpha_n|\ll1\). Throughout this section, $n\in\mathbb{N}$ denotes the winding number, i.e. the image order: a solution with total bending $\alpha(b,t_o)\simeq 2\pi n+\Delta\alpha_n$ corresponds to the $n$-th relativistic image formed by a photon that executes $n$ azimuthal winds near the photon sphere ($n=1$ for one whirl, $n=2$ for two whirls, etc.). Using Eq. \eqref{36} at fixed observer time \(t_o\) and solving for \(b\) near \(b_c(t_o)\) gives the instantaneous sequence of critical impact parameters
\begin{align}
b_n(t_o)&=b_c(t_o)\left[1+s_n(t_o)\right], \nonumber \\
s_n(t_o)&\equiv \exp\,\left(\frac{\bar b(t_o)-2\pi n}{\bar a(t_o)}\right), \label{40}
\end{align}
which is the time-dependent generalization of the standard static SDL hierarchy. The corresponding image angles on the screen are \(\theta_n(t_o)=b_n(t_o)/r_o\), where \(r_o\) is the observer radius introduced in Sec. \ref{subsec2.2}. These \(\theta_n(t_o)\) are the actual angular locations of the \(n\)-th relativistic images on the observer's screen. In particular, each solution \(b_n(t_o)\) of the time-dependent lens equation defines a point on the image plane through the tetrad map Eq. \eqref{12}, and the accumulation point of the sequence \(\theta_n(t_o)\) defines the instantaneous photon-ring radius. Thus, within the present analytic treatment, the observable content in the strong-deflection regime is not limited to the existence of a logarithmic structure in \(\alpha(b,t_o)\): it includes explicit predictions for image positions, their spacing hierarchy, the ring radius, and the associated time-delay modulation.

To expose the imprint of the \((\ell,m)=(2,0)\) QNM, we linearize Eq. \eqref{40} in \(\varepsilon\) using Eq. \eqref{37}. Writing \(\theta_c\equiv b_c/r_o\) and \(s_n^{(0)}\equiv \exp\,\left[(\bar b-2\pi n)/\bar a\right]\), we obtain
\begin{equation}
\theta_n(t_o)=\theta_c\left(1+s_n^{(0)}\right)
+\varepsilon\,\mathrm{Re}\left\{e^{-i\omega t_o}\,\delta\theta_n\right\}
+\mathcal{O}(\varepsilon\,s_n^{(0)}{}^{2}), \label{41}
\end{equation}
with the complex response amplitude
\begin{equation}
\delta\theta_n=\theta_c\,\beta_1\left(1+s_n^{(0)}\right)
+\theta_c\,s_n^{(0)}\,\left(\frac{b_1}{\bar a}-\frac{\bar b-2\pi n}{\bar a^{2}}\,a_1\right). \label{42}
\end{equation}
Eqs. \eqref{41}-\eqref{42} show that all relativistic images wobble coherently at the QNM frequency \(\omega\), with an amplitude that decays geometrically with \(n\) through \(s_n^{(0)}\).
\begin{figure}
    \centering
    \includegraphics[width=0.60\textwidth]{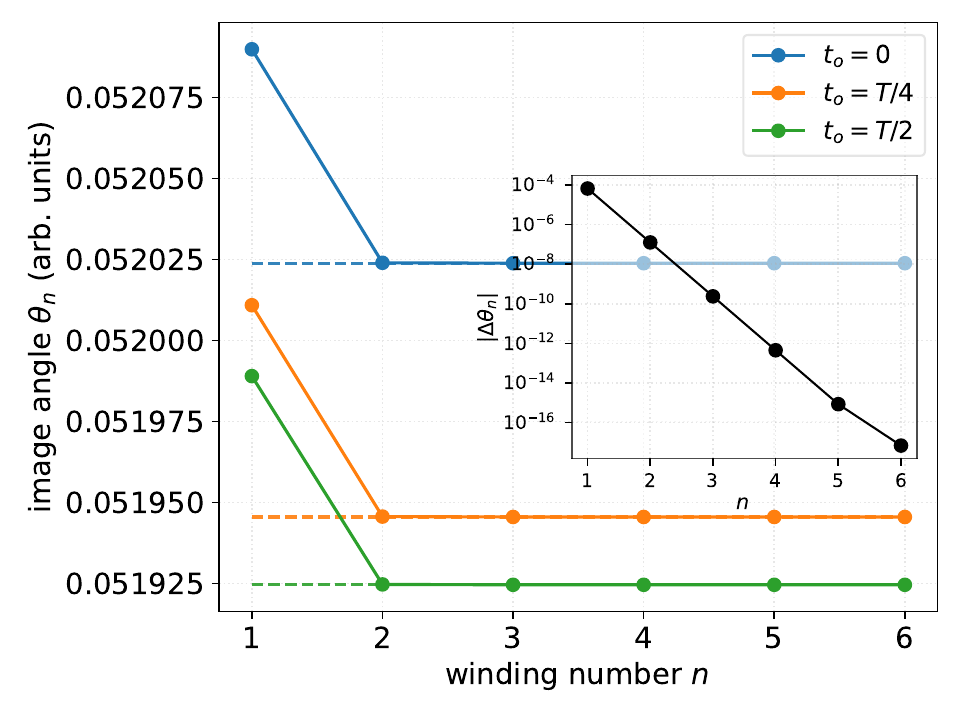}
    \caption{Markers: solid points show the angular positions of the $n$-th relativistic image, $\theta_n(t_o)$; the black dashed line shows the instantaneous photon-ring radius $\theta_{\rm ring}(t_o)=b_c(t_o)/r_o$ [Eq. \eqref{47}]. The rapid convergence $\theta_n \to \theta_{\rm ring}$ with increasing $n$ follows the static SDL spacing $s_n^{(0)}=\exp[(\bar b-2\pi n)/\bar a]$, while the coherent, small-amplitude wobble of all $\theta_n$ with $t_o$ is the QNM-driven modulation predicted by Eqs. \eqref{41}-\eqref{42}. Inset (semilog $y$-axis): the difference $|\Delta\theta_n| \equiv |\theta_n-\theta_{\rm ring}|$ versus $n$ at a fixed $t_o$, highlighting the geometric decay $|\Delta\theta_n|\simeq \theta_c\,s_n^{(0)}$ with baseline slope $\ln q_0=-2\pi$ (Schwarzschild, $\bar a=1$) and a weak, frequency-locked modulation from Eq. \eqref{58}.}
    \label{fig4}
\end{figure}
Figure~\ref{fig4} displays the expected geometric accumulation of images toward the ring, with a phase-locked modulation at frequency $\omega$ inherited from Eq. \eqref{22}. The $\beta_1$ contribution shifts the ring itself, while the $a_1$ and $b_1$ terms imprint the subleading $n$-dependent offsets in Eq. \eqref{42}.

The same near-photon-sphere control that yields Eq. \eqref{36} provides the coordinate travel time for a near-critical ray. Denote by \(T(b,t_o)\) the total coordinate time elapsed between emission and detection for a ray of impact parameter \(b\). By standard SDL analysis, one finds a logarithmic form
\begin{align}
T(b,t_o)&=-\tilde a(t_o)\,\ln\,\left(\frac{b}{b_c(t_o)}-1\right) \nonumber \\
&+\tilde b(t_o)
+ n\,T_{\mathrm{whirl}}(t_o) + \ldots, \label{43}
\end{align}
where \(n\) counts the net azimuthal winds accumulated during the whirl phase, \(T_{\mathrm{whirl}}(t_o)\) is the period of one circular photon orbit measured in Schwarzschild time, and \(\tilde a(t_o),\tilde b(t_o)\) are time-dependent analogues of \(\bar a(t_o),\bar b(t_o)\) governed by the same Born kernel. For Schwarzschild, in the static limit, we have \cite{Ferrari:1984zz,Cardoso:2008bp,Berti:2009kk}
\begin{equation}
T_{\mathrm{whirl}}=\frac{2\pi}{\Omega_{\mathrm{ph}}}, \qquad
\Omega_{\mathrm{ph}}=\frac{1}{3\sqrt{3}M}, \label{44}
\end{equation}
so that successive relativistic images are separated, to leading order, by \(2\pi/\Omega_{\mathrm{ph}}\) in arrival time. Allowing for \(\mathcal{O}(\varepsilon)\) modulations analogous to Eq. \eqref{37}, we write
\begin{align}
\tilde a(t_o)&=\tilde a+\varepsilon\,\mathrm{Re}\left\{e^{-i\omega t_o}\,\tilde a_1\right\},\nonumber \\
\tilde b(t_o)&=\tilde b+\varepsilon\,\mathrm{Re}\left\{e^{-i\omega t_o}\,\tilde b_1\right\},\nonumber \\
T_{\mathrm{whirl}}(t_o)&=T_{\mathrm{whirl}}\left[1+\varepsilon\,\mathrm{Re}\left\{e^{-i\omega t_o}\,\gamma_1\right\}\right], \label{45}
\end{align}
with complex response coefficients \(\tilde a_1,\tilde b_1,\gamma_1\) fixed by the same near-photon-sphere kernel that determines \(a_1,b_1,\beta_1\) in Eq. \eqref{37}. Evaluating Eq. \eqref{43} at \(b=b_n(t_o)\) and forming differences, the inter-image delay becomes
\begin{align}
\Delta T_{n+1,n}(t_o)&\equiv T\left(b_{n+1}(t_o),t_o\right)-T\left(b_n(t_o),t_o\right) \nonumber \\
&= T_{\mathrm{whirl}}(t_o)+\frac{2\pi}{\bar a(t_o)}\,\tilde a(t_o)+\ldots, \label{46}
\end{align}
where the second term arises from the logarithmic piece through \(\ln s_{n+1}-\ln s_n=-2\pi/\bar a(t_o)\). Linearizing Eq. \eqref{46} using Eqs. \eqref{37} and \eqref{45} shows that \(\Delta T_{n+1,n}\) acquires a harmonic modulation \(\propto \mathrm{Re}\,{e^{-i\omega t_o}}\) with an amplitude set by \(\gamma_1,\tilde a_1,a_1\).
\begin{figure}
    \centering
    \includegraphics[width=0.60\textwidth]{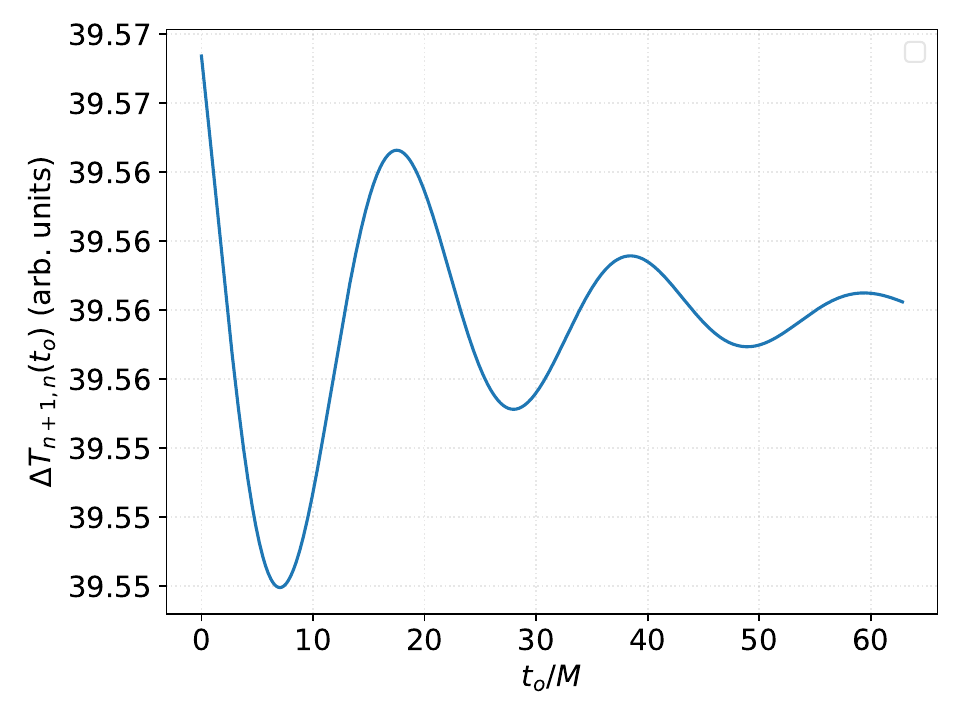}
    \caption{Inter-image time delays $\Delta T_{n+1,n}(t_o)$ from Eq. \eqref{44}. The baseline separation is set by the whirl period $T_{\rm whirl}=2\pi/\Omega_{\rm ph}$ with $\Omega_{\rm ph}=(3\sqrt{3}\,M)^{-1}$, while the weak modulation arises from the time-dependent coefficients in Eq. \eqref{45} and the $\bar a(t_o)$ dependence. The delays oscillate in phase with the spacing modulation $q(t_o)$, reflecting their common origin in the near-photon-sphere kernel.}
    \label{fig6}
\end{figure}
Figure \ref{fig6} makes the phase-locked behavior explicit: variations in $\Delta T_{n+1,n}(t_o)$ track the same QNM phase that controls $q(t_o)$ and $\theta_{\rm ring}(t_o)$. Eliminating $t_o$ between Eqs. \eqref{46} and \eqref{48} yields the differential correlation summarized in Eq. \eqref{60}, further illustrating their common kernel origin.

Finally, the photon ring corresponds to the accumulation of the sequence \(\theta_n(t_o)\) as \(n\to\infty\) \cite{Gralla:2019xty}. From Eq. \eqref{41} we obtain the instantaneous ring radius
\begin{align}
\theta_{\mathrm{ring}}(t_o)&=\lim_{n\to\infty}\theta_n(t_o)
=\frac{b_c(t_o)}{r_o} \nonumber \\
&=\theta_c\left[1+\varepsilon\,\mathrm{Re}\left\{e^{-i\omega t_o}\,\beta_1\right\}\right], \label{47}
\end{align}
so the fractional modulation of the ring radius equals \(\varepsilon\,\mathrm{Re}{e^{-i\omega t_o}\beta_1}\). The exponential spacing of successive relativistic images is governed by
\begin{align}
q(t_o)&\equiv \frac{\theta_{n+1}(t_o)-\theta_{\mathrm{ring}}(t_o)}{\theta_{n}(t_o)-\theta_{\mathrm{ring}}(t_o)}
=\exp\,\left(-\frac{2\pi}{\bar a(t_o)}\right) \nonumber \\
&= e^{-2\pi}\left[1+\varepsilon\,\mathrm{Re}\left\{e^{-i\omega t_o}\,\frac{2\pi\,a_1}{\bar a^{2}}\right\}\right]
+\mathcal{O}(\varepsilon^2), \label{48}
\end{align}
where we used \(\bar a=1\) for Schwarzschild in the static limit. Thus, both the ring radius and the logarithmic spacing inherit the QNM frequency and phase from Eq. \eqref{22}. The same analysis yields the magnifications \(\mu_n(t_o)\), which scale as \(\mu_n\propto q(t_o)^n\) up to a slowly varying prefactor; we refrain from writing the prefactor explicitly because it depends on the source configuration, while the time-dependent exponential factor follows directly from Eq. \eqref{48}. Related visibility-domain signatures and detectability of higher-order rings in realistic accretion models are discussed in \cite{Vincent:2022fwj}; an analytic gap parameter characterization of higher-order rings in spherically symmetric metrics is given in \cite{Aratore:2024bro}.
\begin{figure}
    \centering
    \includegraphics[width=0.60\textwidth]{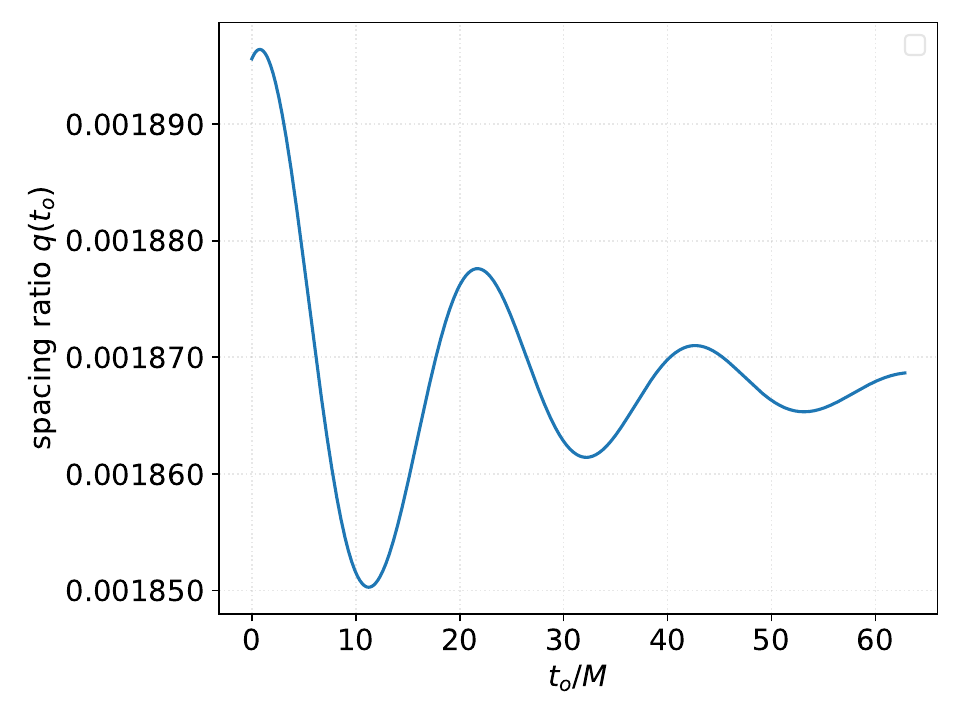}
    \caption{Instantaneous spacing ratio $q(t_o)$ between successive relativistic images. For Schwarzschild, $q_0=e^{-2\pi/\bar a}=e^{-2\pi}$ provides the static baseline, while the small QNM-driven modulation is $\propto \Re\{e^{-i\omega t_o} a_1\}$. Because $q(t_o)$ controls both the inter-image angular spacing and (up to a prefactor) magnification ratios, it supplies a compact strong-field diagnostic tied to the same kernel coefficients as the ring modulation.}
    \label{fig5}
\end{figure}
As shown in Fig. \ref{fig5}, the spacing ratio oscillates weakly around the Schwarzschild value $q_0=e^{-2\pi}$, with phase fixed by the QNM. Together with $\theta_{\rm ring}(t_o)$, this quantity isolates the pair $(a_1,\beta_1)$ that encode the near-photon-sphere response of the Born kernel.

Eqs. \eqref{41}-\eqref{48}, together with the deflection law in Eq. \eqref{36}, provide a self-contained, analytic description of how a single even-parity \(\ell=2,m=0\) QNM coherently imprints on the arrival times, multiplicity, and morphology of strong-field images, culminating in a phase-locked modulation of the photon ring.

\section{Imaging Diagnostics and Continuity Across Regimes} \label{sec5}
We now assemble a single, observer-based description of lensing observables that is valid from the weak field to the time-dependent strong-deflection limit. The central object is the Born line-of-sight map in Eq. \eqref{22}, which yielded the far-field expansion of Eq. \eqref{28} and the strong-field law of Eq. \eqref{36}. Our task is twofold: first, to show that these limits admit a common overlap where they agree term by term once expressed on the screen defined in Sec. \ref{subsec2.2}; second, to translate the matched deflection into gauge-safe imaging diagnostics for centroid motion, multi-image structure, and photon-ring modulations.

\subsection{The unified matching} \label{subsec5.1}
The matching proceeds directly at the level of the master integral in Eq. \eqref{22}. We split the integral at an arbitrary radius \(r_m\) chosen so that \(3M<r_m\ll r_o\), writing
\begin{align}
\mathcal{A}_A(b,\omega)
&=\int_{\lambda_s}^{\lambda(r_m)} d\lambda\,\tilde{\mathcal{K}}_A\,e^{i\omega\tau} \nonumber \\
&+\int_{\lambda(r_m)}^{\lambda_o} d\lambda\,\tilde{\mathcal{K}}_A\,e^{i\omega\tau}
\equiv \mathcal{A}_A^{\text{in}}+\mathcal{A}_A^{\text{out}}, \label{49}
\end{align}
where \(A\in{X,Y}\) and all quantities are evaluated along the background ray labeled by \(b\). The inner piece \(\mathcal{A}_A^{\text{in}}\) is dominated by the whirl segment near \(r_{\mathrm{ph}}=3M\); the outer piece \(\mathcal{A}_A^{\text{out}}\) samples the nearly straight leg through the asymptotic domain. Both contributions depend on \(r_m\), but their sum does not.

For \(b\to b_c^{+}\) at fixed \(r_m\), the inner integral reproduces the logarithmic structure derived in Eq. \eqref{35}, hence
\begin{align}
\delta\alpha_A^{\text{in}}(b,t_o)
&=\mathrm{Re}\biggl\{e^{-i\omega t_o}\,\Bigl[\mathcal{C}_A(\omega)\,\ln\,\left(\frac{b}{b_c}-1\right) \nonumber \\
&+\mathcal{D}_A(\omega,r_m)\Bigr]\biggr\}+\ldots, \label{50}
\end{align}
with \(\mathcal{C}_A\) independent of \(r_m\) and \(\mathcal{D}_A\) carrying the \(r_m\)-dependence required to cancel that of the outer piece. Conversely, for \(M/b\ll1\) at fixed \(r_m\), the outer contribution matches the far-field form in Eq. \eqref{25},
\begin{equation}
\delta\alpha_A^{\text{out}}(b,t_o)
=\mathrm{Re}\left\{e^{-i\omega t_o}\,\frac{c_A(\omega,r_m)}{b}\,\mathcal{F}_A(\omega b)\right\}
+\mathcal{O}\,\left(\frac{M}{b^2}\right), \label{51}
\end{equation}
where \(c_A(\omega,r_m)\) depends on the cutoff but \(\mathcal{F}_A\) does not. Adding Eqs. \eqref{50}-\eqref{51} yields a representation that is uniformly valid for \(b> b_c\).
\begin{figure}
    \centering
    \includegraphics[width=0.60\textwidth]{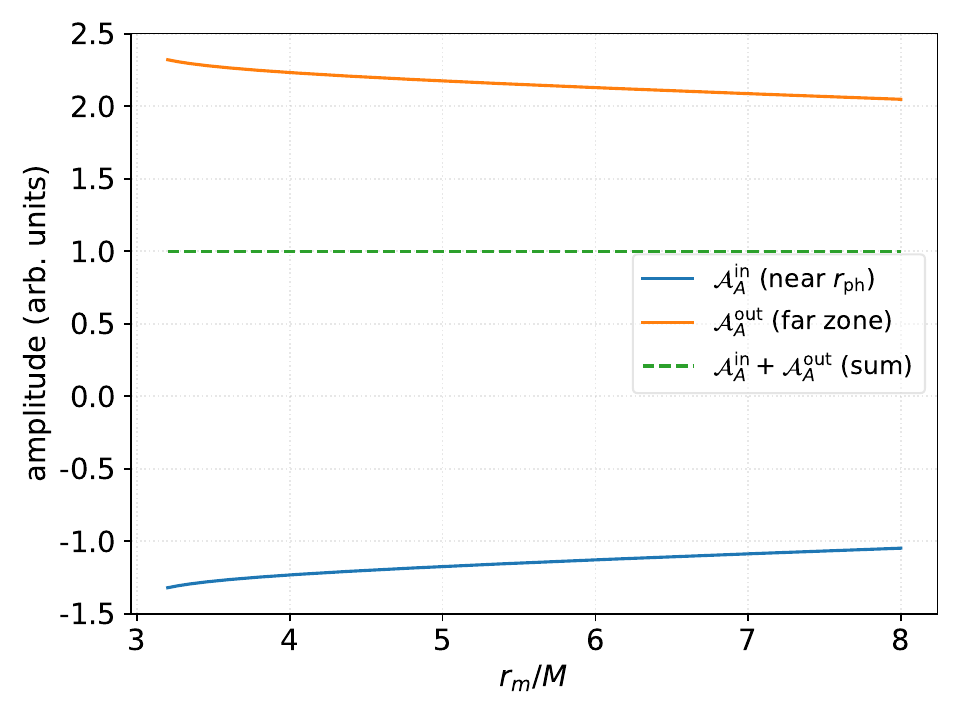}
    \caption{Kernel split and $r_m$ independence of the sum. The master line-of-sight amplitude is decomposed as in Eq. \eqref{49} into the inner (near-photon-sphere) piece and the outer (far-zone) piece. Each contribution depends on the arbitrary split radius $r_m$ and carries complementary logarithmic/finite parts, yet their sum is independent of $r_m$. This illustrates that physical observables derived from the Born kernel are unique and do not depend on the cutoff.}
    \label{fig10}
\end{figure}
Figure \ref{fig10} shows how the $r_m$-dependence cancels between the inner and outer integrals in Eq. \eqref{49}. This cancellation underlies the robustness of the matched coefficients in Eq. \eqref{52} and, ultimately, the overlap relation in Eq. \eqref{53}.

To make contact with the time-dependent SDL coefficients of Eq. \eqref{37}, we expand the outer factor \(\mathcal{F}_A(\omega b)\) for \(b\) close to \(b_c\). Using the standard near-critical relation between the longitudinal coordinate and the azimuthal angle, Eq. \eqref{34}, one finds that \(\mathcal{F}_A\) carries a logarithmic dependence identical to that of the inner piece. The \(r_m\)-dependence cancels between \(\mathcal{D}_A(\omega,r_m)\) and \(c_A(\omega,r_m)\), and the sum can be cast as
\begin{align}
\delta\alpha_A(b,t_o)
&=\mathrm{Re}\biggl\{e^{-i\omega t_o}\Bigl[-a_{1,A}(\omega)\,\ln\,\left(\frac{b}{b_c}-1\right) \nonumber \\
&+\frac{\beta_{1,A}(\omega)}{\frac{b}{b_c}-1}+b_{1,A}(\omega)\Bigr]\biggr\}+\ldots, \label{52}
\end{align}
which is the vectorial version of Eq. \eqref{39} and fixes, component by component, the response coefficients \(a_{1,A},\beta_{1,A},b_{1,A}\) in terms of the same kernel \(\tilde{\mathcal{K}}_A\) used in the far-field calculation.

A practical matching condition follows by selecting an overlap domain where both asymptotics are valid. Let \(b=b_c(1+\Delta)\) with \(0<\Delta\ll1\) and keep \(M/b\ll1\). Expanding Eq. \eqref{28} to first order in \(\Delta\) and equating to Eq. \eqref{52} yields
\begin{align}
\frac{\boldsymbol{\mathcal{A}}(b_c,\omega)}{b_c}
&=\lim_{\Delta\to0^+}\left[
-a_1(\omega)\,\ln\Delta+\frac{\beta_1(\omega)}{\Delta}+b_1(\omega)\right]\,
\hat{\boldsymbol{e}}_b \nonumber \\
&+\boldsymbol{\Xi}(\omega), \label{53}
\end{align}
where \(\boldsymbol{\mathcal{A}}=(c_X\mathcal{F}_X,c_Y\mathcal{F}_Y)\) from Eq. \eqref{28}, \(\hat{\boldsymbol{e}}_b\) is the screen radial unit vector, and \(\boldsymbol{\Xi}(\omega)\) collects finite terms that are independent of the arbitrary split radius \(r_m\). Expanding $A(b,\omega)$ about $b=b_c(1+\Delta)$ generates the $\ln\Delta$ and $1/\Delta$ structures that match the strong-deflection expansion on the inner side, making the regulator independence manifest. Eq. \eqref{53} states that the \(1/b\) Born amplitude evaluated at \(b_c\) encodes, through its singular \(\Delta\)-expansion, the three time-dependent SDL responses in Eq. \eqref{37}. In particular, the residue of the \(1/\Delta\) pole fixes \(\beta_1(\omega)\) and therefore the photon-ring modulation in Eq. \eqref{47}, while the coefficient of \(\ln\Delta\) fixes \(a_1(\omega)\) and hence the spacing modulation in Eq. \eqref{48}.
\begin{figure}
    \centering
    \includegraphics[width=0.60\textwidth]{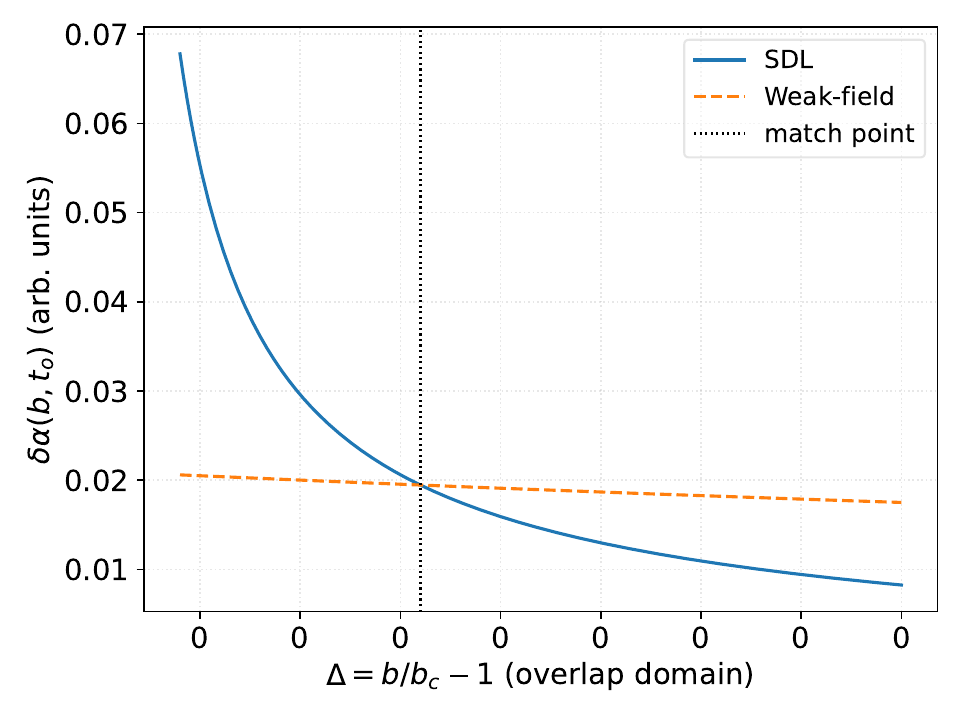}
    \caption{Matching weak and strong limits in the overlap domain. The weak-field Born expression (Eq. \eqref{28}, dashed) and the linearized SDL law (Eq. \eqref{39}, solid) are compared at $b=b_c(1+\Delta)$ for small $\Delta>0$. Around an intermediate reference $\Delta$ the two curves coincide, and they agree within a band across the overlap window, illustrating Eq. \eqref{53}: the near-critical expansion of the $1/b$ Born amplitude encodes the SDL coefficients, in particular the $1/\Delta$ residue that fixes $\beta_1(\omega)$.}
    \label{fig9}
\end{figure}
Figure \ref{fig9} visualizes the content of Eq. \eqref{53}: the same kernel controls both limits, and their expansions meet smoothly for small but finite $\Delta$. The agreement pins down the triplet $(\beta_1,a_1,b_1)$ that governs ring-radius and spacing modulations (Sec. \eqref{sec4}).

Finally, since Eqs. \eqref{22}, \eqref{28}, and \eqref{36} are all constructed using the observer tetrad of Eq. \eqref{8}, the matching preserves the gauge safety guaranteed by Eq. \eqref{15}. The result is a single kernel-based description in which weak-field centroid wobble, intermediate-field image shifts, and strong-field ring modulations are different faces of the same frequency-resolved response.

\subsection{Screen-plane observables: centroid, multi-images, and ring} \label{subsec5.2}
All observables are defined on the screen introduced in Eq. \eqref{12} and inherit their time dependence from the master kernel in Eq. \eqref{22}. We emphasize at the outset that \(\boldsymbol{\alpha}(b,t_o)\) in the present framework is already a screen-plane quantity measured by the asymptotic observer, not an auxiliary coordinate bending angle. Accordingly, the formalism reconstructs image-plane kinematics directly: once a ray family is specified, it yields the screen position of each ray, the centroid of a narrow bundle, the locations of the relativistic images, the photon-ring radius, their logarithmic spacing, the inter-image delays, and the leading hierarchy of magnification ratios. What it does \emph{not} attempt to provide by itself is a full synthetic brightness map, which would require additional astrophysical input such as source emissivity and the complete source-to-screen Jacobian.

We first package the screen coordinates into a single complex angle
\begin{equation}
\zeta \equiv \frac{X+iY}{r_o}, \label{54}
\end{equation}
so that a ray of impact parameter \(b\) arriving at \((X,Y)\) corresponds to
\begin{equation}
\zeta(b,t_o)=\frac{\alpha_X(b,t_o)}{r_o}+i\,\frac{\alpha_Y(b,t_o)}{r_o},
\label{54a}
\end{equation}
with \(\boldsymbol{\alpha}\) given by Eq. \eqref{18}. Equations \eqref{54}-\eqref{54a} therefore provide the direct dictionary between the deflection map and the observer's image plane. For a narrow bundle around \(b\), the centroid angle is the intensity-weighted mean of the complex screen coordinate \(\zeta\),
\begin{equation}
\zeta_c(t_o)=\zeta_{c,0}+\varepsilon\,\mathrm{Re}\left\{e^{-i\omega t_o}\,\mathcal{B}(\omega)\right\}
+\mathcal{O}\,\left(\frac{M}{r_o}\frac{M}{b^2}\right), \label{55}
\end{equation}
which is the complex form of Eq. \eqref{30}. The complex amplitude \(\mathcal{B}(\omega)\) depends on the bundle's impact-parameter distribution through \(\mathcal{A}_A(b,\omega)\) in Eq. \eqref{22}. Since both the phase and modulus of \(\mathcal{B}\) are determined by the same kernel that appears in the strong-field coefficients of Eq. \eqref{37}, the centroid wobble is phase-locked to the modulations of the relativistic images and of the photon ring discussed below. In particular, defining
\begin{equation}
\Phi_{\rm c\,-\,ring}\equiv \arg\mathcal{B}-\arg\beta_1, \label{56}
\end{equation}
the relative phase between the weak-field centroid wobble and the strong-field ring-radius oscillation, Eq. \eqref{47}, is an overlap-domain diagnostic of the common QNM driver.
\begin{figure}
    \centering
    \includegraphics[width=0.60\textwidth]{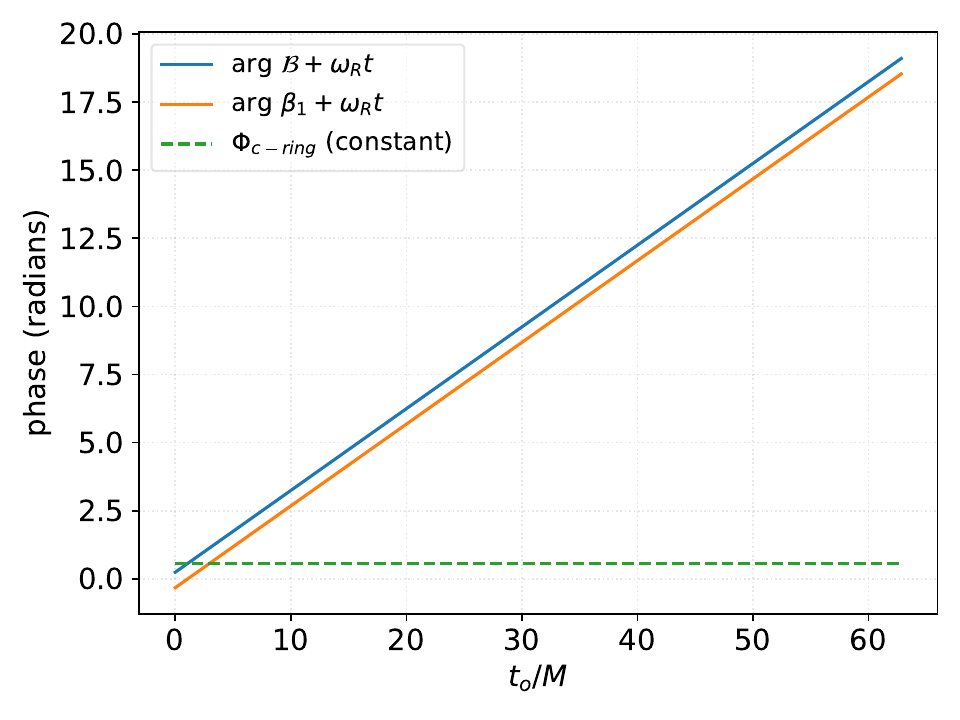}
    \caption{Phase-lock diagnostic across regimes. We show the phases associated with the weak-field centroid wobble (from Eq. \eqref{30}) and the strong-field ring-radius modulation (from Eq. \eqref{47}), together with their difference $\Phi_{\rm c\!-\!ring}$ defined in Eq. \eqref{56}. The two phases advance in parallel at frequency $\omega_R$ while their difference remains constant, demonstrating that weak- and strong-field screen observables are driven by the same QNM kernel.}
    \label{fig7}
\end{figure}
Figure \ref{fig7} exhibits the anticipated phase locking: the centroid and ring phases are parallel lines with a constant offset $\Phi_{\rm c\!-\!ring}$. This provides a direct, gauge-safe confirmation that the Born kernel controls both observables, as argued in Sec. \ref{subsec5.1} and encoded in Eq. \eqref{53}.

Relativistic images are labeled by the winding number \(n\in\mathbb{N}\) and obey Eq. \eqref{41}. Their instantaneous angular radii \(\theta_n(t_o)\) separate naturally into a ring-centered offset,
\begin{align}
\Delta\theta_n(t_o)&\equiv \theta_n(t_o)-\theta_{\rm ring}(t_o) = \theta_c\,s_n^{(0)} \nonumber \\
&+\varepsilon\,\mathrm{Re}\left\{e^{-i\omega t_o}\,\delta\theta_n^{\rm off}\right\}
+\mathcal{O}\,\left(s_n^{(0)}{}^{2}\right), \label{57}
\end{align}
with \(\theta_{\rm ring}(t_o)\) given by Eq. \eqref{47}. The complex offset amplitude,
\begin{equation}
\delta\theta_n^{\rm off}=\theta_c\,s_n^{(0)}\,\left(\frac{b_1}{\bar a}-\frac{\bar b-2\pi n}{\bar a^{2}}\,a_1\right), \label{58}
\end{equation}
is the second term in Eq. \eqref{42}. The geometric decay \(\propto s_n^{(0)}\) ensures that higher-order images accumulate at the ring while remaining coherently modulated at the QNM frequency \(\omega\).

The logarithmic spacing is governed by \(q(t_o)\) in Eq. \eqref{48}. Writing \(q(t_o)=q_0,[1+\varepsilon\,\mathrm{Re}{e^{-i\omega t_o}\,\delta_q}]\) with \(q_0=e^{-2\pi/\bar a}\) and \(\bar a=1\) for Schwarzschild, we have
\begin{equation}
\delta_q=\frac{2\pi\,a_1}{\bar a^{2}}, \label{59}
\end{equation}
so that a measurement of the instantaneous ratio \(\left(\Delta\theta_{n+1}/\Delta\theta_n\right)(t_o)\) yields \(\mathrm{Re}{e^{-i\omega t_o}\delta_q}\) directly. In combination with Eq. \eqref{47}, which determines \(\mathrm{Re}{e^{-i\omega t_o}\beta_1}\), this provides two independent strong-field observables tied to the same kernel coefficients \(a_1\) and \(\beta_1\).

Arrival times follow from the travel-time function \(T(b,t_o)\) in Eq. \eqref{43}. Evaluated at \(b_n(t_o)\) from Eq. \eqref{40}, the inter-image delays \(\Delta T_{n+1,n}(t_o)\) are given by Eq. \eqref{46} and thus oscillate at the QNM frequency through the modulations in Eq. \eqref{45}. Eliminating \(t_o\) between Eqs. \eqref{46} and \eqref{48} gives a purely geometric correlation,
\begin{equation}
\frac{d}{dt_o}\ln q(t_o)
=\frac{2\pi}{\bar a^{2}}\,\frac{d}{dt_o}\bar a(t_o)
=\mathcal{C}_T\,\frac{d}{dt_o}\Delta T_{n+1,n}(t_o)+\mathcal{O}(\varepsilon^2), \label{60}
\end{equation}
where \(\mathcal{C}_T\) is a constant fixed by the static Schwarzschild values of \(\tilde a\) and \(\bar a\). Eq. \eqref{60} expresses the common origin of spacing and delay modulations in the near-photon-sphere kernel and, in our framework, is the time-domain analogue of the matching relation \eqref{53}.

Magnifications \(\mu_n(t_o)\) depend on the source configuration, but their exponential hierarchy is controlled solely by \(q(t_o)\): for a small, fixed source displacement, the leading scaling is
\begin{equation}
\mu_{n+1}(t_o)\simeq q(t_o)\,\mu_n(t_o), \label{61}
\end{equation}
up to a slowly varying prefactor that is independent of \(n\). Using Eq. \eqref{48}, the fractional modulation of \(\mu_n\) at fixed \(n\) is therefore \(\delta\mu_n/\mu_n\simeq \mathrm{Re}{e^{-i\omega t_o}\delta_q}\), while at fixed \(t_o\) the ratio \(\mu_{n+1}/\mu_n\) isolates \(q(t_o)\). Together with Eqs. \eqref{47}-\eqref{48}, this yields a closed set of strong-field observables \({\theta_{\rm ring},q,\mu_{n+1}/\mu_n}\) that are all phase-locked to \(\omega\). These relations also make clear the scope of the present framework. The quantities \(\zeta_c(t_o)\), \(\theta_n(t_o)\), \(\theta_{\rm ring}(t_o)\), \(q(t_o)\), \(\Delta T_{n+1,n}(t_o)\), and \(\mu_{n+1}/\mu_n\) are genuine image-plane or timing observables because they are constructed entirely from the observer tetrad and the screen map. By contrast, a full surface-brightness image or a detailed visibility model would additionally require a source emissivity prescription and the complete lens mapping from source coordinates to the observer screen. Our emphasis here is on the universal, source-independent kinematic imprint of the QNM on the screen observables themselves.

Finally, the weak- and strong-field diagnostics can be combined without reference to the arbitrary matching radius used in Eq. \eqref{49}. A convenient pair is
\begin{equation}
\left(\zeta_c(t_o)-\zeta_{c,0}\right)\quad\text{and}\quad
\theta_{\rm ring}(t_o)-\theta_c, \label{62}
\end{equation}
whose complex phases are \(\arg\mathcal{B}\) and \(\arg\beta_1\), respectively. The constancy of \(\Phi_{\rm c\,-\,ring}\) in Eq. \eqref{56} across the overlap domain provides a direct test of continuity from Eq. \eqref{28} to Eq. \eqref{36}. Because all quantities in Eqs. \eqref{55}-\eqref{62} are defined through the tetrad map Eq. \eqref{12}, the gauge-independence argument of Eq. \eqref{15} applies verbatim: centroid motion, inter-image spacing, ring radius, and magnification ratios are bona fide $\mathcal{O}(\varepsilon)$ observables tied to a common first-order response encoded in the Born kernel \eqref{22}.

\section{Conclusion} \label{sec6}
We have constructed an analytic description that connects weak-field lensing to a time-dependent strong-deflection limit for light propagating in a perturbed Schwarzschild spacetime. The backbone of our construction is the line-of-sight Born map in Eq. \eqref{22}, derived from the first-order response of null geodesics to an axisymmetric even-parity \((\ell,m)=(2,0)\) QNM perturbation. By evaluating the same kernel on background rays we obtained, on one hand, the far-field expansion of Eq. \eqref{28}, which produces a ringdown-frequency centroid wobble on the observer's screen, and, on the other hand, the near-photon-sphere expansion of Eq. \eqref{36}, which yields a time-dependent generalization of the logarithmic strong-deflection law. The linearized form in Eq. \eqref{39} exposes the dominance of the critical-scale modulation as \(b\to b_c^{+}\). This single-kernel perspective makes explicit that weak-field and strong-field observables are phase-locked to the same QNM frequency and are related by controlled asymptotics.

The imaging layer built on the observer tetrad of Eq. \eqref{8} renders all diagnostics gauge-safe at \(\mathcal{O}(\varepsilon)\). In the weak field, the centroid \(\boldsymbol{\theta}_c(t_o)\) follows Eq. \eqref{30} and provides a clean harmonic signature. In the strong field, the hierarchy of relativistic images \(\theta_n(t_o)\) in Eq. \eqref{41} and the photon-ring radius \(\theta_{\mathrm{ring}}(t_o)\) in Eq. \eqref{47} exhibit coherent, frequency-resolved modulations. The logarithmic spacing \(q(t_o)\) in Eq. \eqref{48} and the inter-image delays in Eq. \eqref{46} trace the same near-photon-sphere kernel that controls the modulation of \(b_c(t_o)\) via Eq. \eqref{37}. The matching procedure of Sec. \ref{subsec5.1}, summarized by Eq. \eqref{53}, makes the continuity across regimes explicit: the residue that governs the photon-ring modulation and the coefficient that governs spacing variations are encoded in the near-critical expansion of the very \(1/b\) amplitude that drives the centroid wobble. For a global, quasi-normal oscillation on a black-hole background, the leading Born imprint scales as $1/b$; for localized radiative sources, near-zone terms dominate and yield net $b^{-3}$ scaling after intermediate-wave zone cancellations.

Several extensions can be pursued within the same analytic setup. One may include additional even-parity multipoles, or odd-parity counterparts, by incorporating the corresponding gauge-invariant master fields into the Born kernel; the derivations follow the same steps that lead to Eqs. \eqref{22}, \eqref{28}, and \eqref{36}. Although we have focused on Schwarzschild, the strategy can be generalized to slowly rotating backgrounds by treating the spin as a controlled perturbation, replacing the Zerilli sector with its Teukolsky analogue while keeping the observer-plane definitions fixed. The present construction also motivates a systematic study of phase relations among weak-field wobble, spacing, and ring modulations, such as \(\Phi_{\rm c-ring}\) in Eq.~\eqref{56}, as kernel-level diagnostics that are insensitive to source morphology. Finally, the appendices collect the reconstruction and kernel-evaluation steps used here and can be adapted to higher-order corrections when these remain under analytic control.

To end, expressing the deflection and the associated imaging observables in terms of a single time-dependent Born representation makes the weak-to-strong continuity transparent and keeps the gauge control explicit. The resulting relations among centroid motion, multi-image structure, and photon-ring modulations clarify how ringdown-scale perturbations enter screen observables around Schwarzschild black holes.

\appendix
\section{Zerilli framework and reconstruction to \texorpdfstring{\(h_{\mu\nu}\)}{}} \label{appA}
In this appendix, we collect the even-parity RWZ machinery used in Secs. \ref{sec2}-\ref{sec5}. We keep the background Schwarzschild metric \(g^{(0)}_{\mu\nu}\) and notation of Eq. \eqref{1}, the perturbative split of Eq. \eqref{2}, and the Zerilli master equation and potential of Eqs. \eqref{5}-\eqref{6}, together with the QNM boundary conditions in Eq. \eqref{7}. Our goal is twofold: to define a gauge-invariant even-parity master field \(\Psi_Z\) appropriate for \((\ell,m)=(2,0)\), and to reconstruct the RWZ-gauge metric amplitudes \(H_0,H_1,H_2,K\) and the combination \(h_{\alpha\beta}p^\alpha p^\beta\) that appears in the Born kernel of Eqs. \eqref{19}-\eqref{22}.

\emph{Harmonic decomposition and RWZ gauge:} We expand first-order even-parity perturbations in scalar spherical harmonics \(Y_{\ell m}(\theta,\phi)\),
\begin{align}
h_{tt} &= f\,H_0^{\ell m}(t,r)\,Y_{\ell m},\nonumber \\
h_{tr} &= H_1^{\ell m}(t,r)\,Y_{\ell m},\nonumber \\
h_{rr} &= f^{-1}H_2^{\ell m}(t,r)\,Y_{\ell m}, \nonumber \\
h_{tA} &= h_{rA} = 0,\nonumber \\
h_{AB} &= r^2K^{\ell m}(t,r)\,\Omega_{AB}\,Y_{\ell m}, \label{A.1}
\end{align}
with \(A,B\in{\theta,\phi}\), \(\Omega_{AB}=\mathrm{diag}(1,\sin^2\theta)\), and \(f=1-2M/r\) as in Eq. \eqref{1}. The RWZ gauge conditions set the even-parity vector and tensor harmonics \((j_{a},G)\) to zero. For axisymmetry ($m=0$) we have $Y_{\ell 0}(\theta)=\sqrt{\frac{2\ell+1}{4\pi}}\,P_\ell(\cos\theta)$. All amplitudes $H_0,H_1,H_2,K$ used in the main text (Secs. \ref{sec2}-\ref{sec5}) are defined in the $P_\ell$ basis. When needed, $Y$-basis amplitudes and $P$-basis amplitudes are related by $H_a^{(Y)}=\sqrt{\tfrac{2\ell+1}{4\pi}}\,H_a^{(P)}$ for $a\in\{0,1,2,K\}$. It is also convenient to adopt the tortoise coordinate \(r_*\) of Eq. \eqref{4}, so that the wave operator acting on master fields takes the flat-space form in \((t,r_*)\).

\emph{Gauge-invariant Zerilli-Moncrief master field:} Gauge invariants for the even-parity sector can be built \`a la Moncrief. A convenient choice is the Zerilli-Moncrief master field \(\Psi_Z(t,r)\) defined by
\begin{align}
\Psi_Z &= \frac{r}{\lambda r+3M}\Bigg[
K+\frac{r-2M}{\lambda r+3M}\left(H_2 - r\,\partial_r K\right)\Bigg], \nonumber \\
\lambda&\equiv\frac{(\ell-1)(\ell+2)}{2}=2. \label{A.2}
\end{align}
By construction, \(\Psi_Z\) is invariant under linearized even-parity gauge transformations [cf. Eq. \eqref{14}]. In vacuum it satisfies the Zerilli equation already quoted in Eq. \eqref{5},
\begin{equation}
\left(-\partial_t^2+\partial_{r_*}^2-V_Z(r)\right)\Psi_Z=0, \label{A.3}
\end{equation}
with \(V_Z\) given in Eq. \eqref{6}. For QNMs, we impose the homogeneous boundary conditions of Eq. \eqref{7}. In the frequency domain, \(\Psi_Z(t,r)=\psi(r)\,e^{-i\omega t}\) with complex \(\omega=\omega_R+i\omega_I\), \(\omega_I<0\).

A normalization convention is needed to separate the physical smallness of the perturbation from the shape of \(\Psi_Z\). Throughout we take the outgoing amplitude of \(\Psi_Z\) at large \(r\) to be unity and collect the overall smallness in the book-keeping parameter \(\varepsilon\) of Eq. \eqref{2}; this is the convention used in Secs. \ref{sec3}-\ref{sec5}.

\emph{Reconstruction of \(H_0,H_1,H_2,K\) from \(\Psi_Z\):} In vacuum, the Einstein equations relate \((H_0,H_1,H_2,K)\) algebraically to \(\Psi_Z\) and its first derivatives. One convenient RWZ-gauge reconstruction, obtained by inverting Eq. \eqref{A.2} and using the field equations, reads
\begin{align}
K &= \frac{\mathcal{N}_K(r)}{r^2(\lambda r+3M)}\,\Psi_Z
+\frac{f}{r}\,\partial_r\Psi_Z, \nonumber \\
H_2 &= \frac{\mathcal{N}_{2}(r)}{r^2 f(\lambda r+3M)}\,\Psi_Z
+\frac{\mathcal{D}_{2}(r)}{r f}\,\partial_r\Psi_Z, \nonumber \\
H_0 &= H_2, \nonumber \\
H_1 &= -\frac{i\omega\,\mathcal{N}_{1}(r)}{f(\lambda r+3M)}\,\Psi_Z
+ i\omega\, r, \partial_r\Psi_Z , \label{A.4}
\end{align}
where \(\omega\) is the Fourier frequency introduced above and the radial polynomials are
\begin{align}
\mathcal{N}_K&=\lambda(\lambda+1)r^2+3\lambda M r+6 M^2, \nonumber \\
\mathcal{N}_{2}&=\lambda(\lambda+1) r^2 -3\lambda M r - 6 M^2, \nonumber \\
\mathcal{D}_{2}&=\lambda r - 3M, \nonumber \\
\mathcal{N}_{1}&=\lambda r^2+3\lambda M r + 6 M^2. \label{A.5}
\end{align}
For Schwarzschild with \(\ell=2\) one has \(\lambda=2\). Eq. A.4 is written in a form that makes the large-\(r\) behavior transparent: \(K\sim r^{-1}\), \(H_2\sim r^{-1}\), \(H_1\sim r^{-1}\) once \(\Psi_Z\sim r^{-1}e^{-i\omega(t-r_*)}\).

We remark that (i) The equality \(H_0=H_2\) follows from the vacuum field equations in RWZ gauge; (ii) Other algebraically equivalent reconstructions exist; any such choice leads to the same gauge-invariant observables; (iii) The particular coefficients in Eq. \eqref{A.5} are tailored to the definition in Eq. \eqref{A.2}. If a different master variable is adopted, the polynomials change accordingly without affecting physical content.

\emph{Asymptotics and horizon behavior:} From Eq. \eqref{7}, \(\Psi_Z\) behaves as an outgoing wave at infinity and as an ingoing wave at the horizon,
\begin{equation}
\Psi_Z \sim \begin{cases}
A_\infty\, e^{-i\omega(t-r_*)}, & r\to\infty, \\
A_H\, e^{-i\omega(t+r_*)}, & r\to 2M,
\end{cases} \label{A.6}
\end{equation}
with \(A_\infty=1\) by our convention and \(A_H\) determined by the solution of Eq. \eqref{A.3}. Substituting \(A.6\) into \(A.4\) and using \(r_*\sim r+2M\ln(r/2M-1)\), we find
\begin{align}
H_{0,1,2}\,K &= \mathcal{O}(r^{-1})\,e^{-i\omega(t-r)} \quad \text{for} \quad  (r\to\infty),
 \nonumber \\
H_{0,1,2},K &= \mathcal{O}(1)\,e^{-i\omega(t+r_*)} \quad \text{for} \quad (r\to 2M), \label{A.7}
\end{align}
which ensures that the perturbed metric decays as \(r^{-1}\) at infinity and is regular on the future horizon in ingoing Eddington-Finkelstein coordinates.

\emph{The contraction \(h_{\alpha\beta}p^\alpha p^\beta\) along a background ray:} The Born kernel in Eqs. \eqref{19}-\eqref{22} involves \(\partial_\mu h_{\alpha\beta}\,p_0^\alpha p_0^\beta\) evaluated on the Schwarzschild null geodesic \(x_0^\mu(\lambda,b)\). Using Eq. \eqref{A.1} with \(m=0\) and the axisymmetric harmonic \(Y_{20}(\theta)=\sqrt{5/(4\pi)}\,P_2(\cos\theta)\), we obtain
\begin{align}
h_{\alpha\beta}p_0^\alpha p_0^\beta
&= Y_{20}(\theta)\,\Bigl[
f H_0\, (p_0^t)^2 + 2 H_1\, p_0^t p_0^r \nonumber \\ 
&+ f^{-1}H_2\, (p_0^r)^2
+ r^2 K \,\Omega_{AB}, \hat p_0^A \hat p_0^B
\Bigr], \label{A.8}
\end{align}
where \(\hat p_0^A=(p_0^\theta\,p_0^\phi)\) are the angular components of the background momentum and \(\Omega_{AB}\,\hat p_0^A\hat p_0^B=(p_0^\theta)^2+\sin^2\theta,(p_0^\phi)^2=L^2/r^4\) with \(L\) the conserved angular momentum of the background geodesic [cf. Eq. \eqref{13}]. Substituting Eq. \eqref{A.4} into Eq. \eqref{A.8} and simplifying gives the compact representation
\begin{equation}
h_{\alpha\beta}p_0^\alpha p_0^\beta
= Y_{20}(\theta)\left[\mathcal{P}_0(r,E,L)\,\Psi_Z
+\mathcal{P}_1(r,E,L)\,\partial_r\Psi_Z\right], \label{A.9}
\end{equation}
with
\begin{align}
\mathcal{P}_0(r,E,L) &=
\frac{\mathcal{A}_0(r)}{r^2(\lambda r+3M)}\,(E^2 - f L^2/r^2) \nonumber \\
&- \frac{2 i\omega\,\mathcal{N}_{1}(r)}{f(\lambda r+3M)}\,E\,\sqrt{E^2-fL^2/r^2},
\mathcal{P}_1(r,E,L) \notag \\&=
\frac{f}{r}\,(E^2 - f L^2/r^2) + 2 i\omega r\,E\,\sqrt{E^2-fL^2/r^2}, \label{A.10}
\end{align}
where \(\mathcal{A}_0(r)\equiv \mathcal{N}_K(r)+\mathcal{N}_2(r)\) and we have expressed the radial momentum as \(p_0^r=\pm \sqrt{E^2-fL^2/r^2}\) with \(E=-p_{0t}\) the conserved energy (the sign flips at turning points). Eqs. \eqref{A.9}-\eqref{A.10} make explicit that only \(\Psi_Z\) and \(\partial_r\Psi_Z\) enter the kernel, and that the dependence on the orbital constants \((E,L)\) factorizes algebraically.

Taking a derivative and using the background geodesic equations, the source term in Eq. \eqref{17} can be written in the equivalent forms
\begin{align}
\partial_\mu\,\left(h_{\alpha\beta}p_0^\alpha p_0^\beta\right)\,dx_0^\mu
&= \left(\partial_t - \dot r_0\,\partial_r\right)
\left[h_{\alpha\beta}p_0^\alpha p_0^\beta\right]\,d\lambda \nonumber \\
&= \left(-i\omega + \frac{dr_*}{d\lambda}\,\partial_{r_*}\right)
\left[h_{\alpha\beta}p_0^\alpha p_0^\beta\right]\,d\lambda\, \label{A.11}
\end{align}
which is useful for casting the line-of-sight integrals \eqref{19}-\eqref{22} in terms of the tortoise coordinate \(r_*\) and the delay \(\tau\) used in Eq. \eqref{22}.

\emph{Large-(r) limit and normalization check:} With the outgoing normalization \(A_\infty=1\) in Eq. \eqref{A.6}, the asymptotic behavior of Eq. \eqref{A.4} implies
\begin{align}
h_{\alpha\beta}p_0^\alpha p_0^\beta
&\sim \frac{Y_{20}(\theta)}{r}\,e^{-i\omega(t-r)}\, \nonumber \\
&\times \left[\mathfrak{c}_0(\omega,E,L) +\mathfrak{c}_1(\omega,E,L)\,\mathcal{O}(1/r)\right], \label{A.12}
\end{align}
for some algebraic coefficients \(\mathfrak{c}_{0,1}\). Substitution of Eq. \eqref{A.12} into the kernel Eq. \eqref{20} reproduces the far-zone scaling used in Sec. \ref{subsec3.2} and leads directly to the \(1/b\) dependence in Eq. \eqref{25}, providing a consistency check of the normalization choices above.

\emph{Near-photon-sphere reduction:} Near the photon sphere \(r_{\mathrm{ph}}=3M\) [Eq. \eqref{31}], the master equation in Eq. \eqref{A.3} can be approximated by a Schr\"odinger problem with an inverted harmonic potential around the unstable maximum of \(V_Z\). Expanding \(\Psi_Z\) and its derivative in that region and inserting into Eq. \eqref{A.4} and then (A.8) yields
\begin{align}
h_{\alpha\beta}p_0^\alpha p_0^\beta
&\simeq Y_{20}(\theta)\Bigl[ \mathfrak{C}_0(\omega,E,L)\,\Psi_Z \nonumber \\
&+\mathfrak{C}_1(\omega,E,L)\,\partial_{r}\Psi_Z\Bigr]_{r=3M}
+\ldots, \label{A.13}
\end{align}
with constants \(\mathfrak{C}_{0,1}\) determined by the polynomials in Eq. \eqref{A.5}. This reduction underlies the logarithmic structures extracted in Sec. \ref{sec4}, culminating in Eqs. \eqref{35}-\eqref{39}.

Eqs. \eqref{A.2}-\eqref{A.5} provide a self-contained RWZ-gauge reconstruction from the Zerilli-Moncrief master field to the metric amplitudes \((H_0,H_1,H_2,K)\). The contraction formulas in Eqs. \eqref{A.8}-\eqref{A.11} deliver the precise input required by the Born kernel of Eq. \eqref{20}, and their far-zone and near-photon-sphere limits, Eqs. \eqref{A.12}-\eqref{A.13}, are the only ingredients needed to derive the weak-field expansion (Sec. \ref{subsec3.2}) and the time-dependent SDL laws (Sec. \ref{sec4}). All steps are consistent with the QNM boundary conditions in Eq. \eqref{7}, and all observable statements inherit gauge safety from the use of \(\Psi_Z\) and the observer tetrad defined in Eq. \eqref{8}.

\section{Derivation of the master Born kernel} \label{appB}
Our goal is to derive the line-of-sight expression that underlies Eqs. \eqref{18}-\eqref{22} for the screen-plane deflection produced by a first-order metric perturbation \(h_{\mu\nu}\) on the Schwarzschild background of Eq. \eqref{1}. We keep the perturbative split of Eq. \eqref{2}, use the observer tetrad of Eq. \eqref{8}, and map momenta to screen angles via Eq. \eqref{12}. Throughout, we follow a Born, or single-scattering, treatment: the null ray follows the unperturbed geodesic \(x_0^\mu(\lambda,b)\) while its momentum \(p^\mu\) is corrected at \(\mathcal{O}(\varepsilon)\).

\emph{From Hamilton's equations to a line-of-sight integral:} Start from the Hamiltonian for null motion
\begin{equation}
H(x,p)=\frac{1}{2}\,g^{\mu\nu}(x)\,p_\mu p_\nu=0, \label{B.1}
\end{equation}
and expand \(g^{\mu\nu}=g^{(0)\mu\nu}-\varepsilon\,h^{\mu\nu}+\mathcal{O}(\varepsilon^2)\), where indices on \(h\) are raised with \(g^{(0)\mu\nu}\). Let \(x^\mu=x_0^\mu+\varepsilon\,\delta x^\mu\) and \(p_\mu=p_{0\mu}+\varepsilon\,\delta p_\mu\). To first order,
\begin{equation}
\dot x_0^\mu=\frac{\partial H_0}{\partial p_{0\mu}}=g^{(0)\mu\nu}p_{0\nu}\equiv p_0^\mu,\qquad
\dot p_{0\mu}=-\frac{\partial H_0}{\partial x_0^\mu}=0, \label{B.2}
\end{equation}
along $x_0^\mu$, where a dot is \(d/d\lambda\) and \(H_0=\frac{1}{2} g^{(0)\mu\nu}p_{0\mu}p_{0\nu}=0\). The first-order Hamilton equation for \(\delta p_\mu\) is
\begin{equation}
\dot{\delta p}_\mu = -\frac{\partial}{\partial x^\mu}\,\left(\frac{1}{2}\,h^{\alpha\beta}p_{0\alpha}p_{0\beta}\right)
= -\frac{1}{2}\,\partial_\mu h_{\alpha\beta}\,p_0^\alpha p_0^\beta, \label{B.3}
\end{equation}
which integrates to Eq. \eqref{17} once evaluated between source \(\lambda_s\) and observer \(\lambda_o\):
\begin{equation}
\delta p_\mu(\lambda_o,b)=-\frac{1}{2}\int_{\lambda_s}^{\lambda_o}\,d\lambda\,\partial_\mu h_{\alpha\beta}\left(x_0(\lambda)\right)\,p_0^\alpha(\lambda)\,p_0^\beta(\lambda). \label{B.4}
\end{equation}
Eq. \eqref{B.4} is the covariant Born response on a fixed background path. At this stage, the remaining steps are purely kinematical. First, \(\delta p_\mu\) is projected onto the observer tetrad at \(x_o^\mu\). Second, the screen map Eq. \eqref{12} is varied at fixed observer event, which converts \(\delta p^{\hat a}\) into the image-plane deflection \(\delta\alpha_A\). Third, substituting Eq. \eqref{B.4} into that projector relation yields the line-of-sight kernel form used in the main text. We spell out these steps below so that Eq. \eqref{20} can be verified directly from the Hamiltonian starting point.

Two comments are useful. First, the force driving \(\delta p_\mu\) can equivalently be written in terms of the perturbed Christoffel symbols, since
\begin{align}
\partial_\mu h_{\alpha\beta}\,p_0^\alpha p_0^\beta
=2\,\delta\Gamma_{\mu\alpha\beta}\,p_0^\alpha p_0^\beta
+\partial_\lambda\,\left(h_{\mu\nu}p_0^\nu\right), \nonumber \\
\delta\Gamma_{\mu\alpha\beta}=\frac{1}{2}\,\left(\nabla_\alpha h_{\mu\beta}
+\nabla_\beta h_{\mu\alpha}-\nabla_\mu h_{\alpha\beta}\right), \label{B.5}
\end{align}
so that integrating by parts turns \(B.4\) into a \(\delta\Gamma\)-driven integral plus boundary terms. Second, the contraction \(h_{\alpha\beta}p_0^\alpha p_0^\beta\) used below is given in \eqref{A.8}-\eqref{A.11} in terms of the Zerilli master field \(\Psi_Z\).

\emph{Projection to the observer's screen:} Let \(e^{\mu}{}_{\hat{a}}(x_o)\) be the tetrad of Eq. \eqref{8} at the observer event \(x_o^\mu\). The measured momentum components are
\begin{equation}
p^{\hat{a}}=e^{\hat{a}}{}_{\mu}p^\mu=p_0^{\hat{a}}+\varepsilon\,\delta p^{\hat{a}},\qquad
\delta p^{\hat{a}}=e^{\hat{a}}{}_{\mu}\,\delta p^\mu, \label{B.6}
\end{equation}
to first order, since the \(\mathcal{O}(\varepsilon)\) tetrad correction only induces \(\mathcal{O}(\varepsilon^2)\) changes in \(p^{\hat{a}}\) when combined with \(\delta p^\mu\); the \(\mathcal{O}(\varepsilon)\) orthonormality is enforced in Eq. \eqref{9}. At fixed observer event \(x_o^\mu\), the screen basis is \((\hat{X},\hat{Y})=(\hat{\phi},-\hat{\theta})\), and the mapping to angles uses Eq. \eqref{12}:
\begin{align}
X&=-\frac{r_o\,p^{\hat{\phi}}}{p^{\hat{0}}},\qquad
Y=\frac{r_o\,p^{\hat{\theta}}}{p^{\hat{0}}},\nonumber \\
\alpha_A&\equiv \frac{A}{r_o}=\frac{p^{\widehat{A}}}{p^{\hat{0}}},\quad A\in{X,Y}. \label{B.7}
\end{align}
Varying Eq. \eqref{B.7} and keeping terms linear in \(\varepsilon\) gives
\begin{equation}
\delta\alpha_A=\frac{\delta p^{\widehat{A}}}{p_0^{\hat{0}}}-\frac{p_0^{\widehat{A}}}{(p_0^{\hat{0}})^2}\,\delta p^{\hat{0}}
=\frac{1}{p_0^{\hat{0}}}\,\Pi_{A}{}^{\ \hat{b}}\,\delta p_{\hat{b}}, \label{B.8}
\end{equation}
where \(\Pi_{A}{}^{\ \hat{b}}\) is the projector onto the screen directions at fixed \(x_o^\mu\),
\begin{equation}
\Pi_{A}{}^{\ \hat{b}}=\delta_{A}{}^{\ \hat{b}}-\alpha_{0,A}\,\delta^{\hat{b}}{}_{\hat{0}}\,,
\qquad \alpha_{0,A}=\frac{p_0^{\widehat{A}}}{p_0^{\hat{0}}}, \label{B.9}
\end{equation}
and hatted indices are raised/lowered with \(\eta_{\hat{a}\hat{b}}\). Using \(\delta p_{\hat{b}}=e_{\hat{b}}{}^{\mu}\,\delta p_\mu\) and substituting Eq. \eqref{B.4} then yields
\begin{align}
\delta\alpha_A(b,t_o)&= -\frac{1}{2\,p_0^{\hat{0}}(\lambda_o)}
\int_{\lambda_s}^{\lambda_o}\,d\lambda\,
\Pi_{A}{}^{\ \hat{b}}\,e_{\hat{b}}{}^{\ \mu}(x_o)\, \nonumber \\
&\times \partial_\mu h_{\alpha\beta}\left(x_0(\lambda)\right)\,p_0^\alpha(\lambda)p_0^\beta(\lambda), \label{B.10}
\end{align}
Equation \eqref{B.10} is the precise intermediate relation between the momentum perturbation and the image-plane response: it shows that the deflection is obtained by first projecting \(\delta p_\mu\) to the observer frame and then varying the screen map, with no additional dynamical assumption beyond the Born approximation already used in Eq. \eqref{B.4}. Furthermore, Eq. \eqref{B.10} is Eq. \eqref{19} with the explicit kernel
\begin{align}
\mathcal{K}_A\,\left[x_0(\lambda),p_0(\lambda)\right]
&=-\frac{1}{2\,p_0^{\hat{0}}(\lambda_o)}\,
\Pi_{A}{}^{\ \hat{b}}\,e_{\hat{b}}{}^{\ \mu}(x_o)\, \nonumber \\
&\times \partial_\mu h_{\alpha\beta}\left(x_0(\lambda)\right)\,p_0^\alpha p_0^\beta.
\label{B.11}
\end{align}
Eq. \eqref{B.11} is identical to Eq. \eqref{20} once we identify the tetrad time leg \(e^{\nu}{}_{\hat{0}}\) and note that \(\Pi_{A}{}^{\ \mu}= \Pi_{A}{}^{\ \hat{b}} e_{\hat{b}}{}^{\ \mu}\).

\emph{Equivalent kernel forms and boundary terms:} Using the identity Eq. \eqref{B.5} and integrating by parts, Eq. \eqref{B.10} becomes
\begin{equation}
\delta\alpha_A=\int_{\lambda_s}^{\lambda_o}\,d\lambda\,
\mathcal{K}^{(\Gamma)}_{A}
+ \mathcal{B}_A\Big|_{\lambda_s}^{\lambda_o}, \label{B.12}
\end{equation}
with
\begin{equation}
\mathcal{K}^{(\Gamma)}_{A} =
-\frac{1}{p_0^{\hat{0}}(\lambda_o)}\,
\Pi_{A}{}^{\ \hat{b}}\,e_{\hat{b}}{}^{\ \mu}(x_o)\,
\delta\Gamma_{\mu\alpha\beta}\left(x_0(\lambda)\right)\,p_0^\alpha p_0^\beta, \notag
\end{equation}
\begin{equation}
\mathcal{B}_A=\frac{1}{2\,p_0^{\hat{0}}(\lambda_o)}\,
\Pi_{A}{}^{\ \hat{b}}\,e_{\hat{b}}{}^{\ \mu}(x_o)\,
h_{\mu\nu}\left(x_0(\lambda)\right)\,p_0^\nu. \label{B.13}
\end{equation}
The boundary term \(\mathcal{B}_A\) vanishes under our asymptotic conditions: at the observer, the screen projector eliminates any \(\hat{0}\) contamination and \(h_{\mu\nu}(x_o)=\mathcal{O}(r_o^{-1})\) [Eq. \eqref{A.7}]; at the source, we assume compact support or sufficiently rapid falloff. Hence, all kernel representatives obtained by redistributing derivatives are equivalent at \(\mathcal{O}(\varepsilon)\).

A further useful form follows by changing variables from affine parameter \(\lambda\) to the tortoise coordinate \(r_*\) or to the delay \(\tau\) of Eq. \eqref{22}. Using Eq. \eqref{A.11},
\begin{equation}
\delta\alpha_A=\int_{\tau=0}^{\infty}\,d\tau\,
\widetilde{\mathcal{K}}_A\left[r(\tau),\theta(\tau)\right]\,
e^{i\omega \tau} \times e^{-i\omega t_o} + \text{c.c.}, \label{B.14}
\end{equation}
which is the factorized form quoted in Eq. \eqref{22} once the QNM time dependence in Eq. \eqref{21} is inserted.

Examining the gauge behavior at \(\mathcal{O}(\varepsilon)\) under a linearized diffeomorphism generated by \(\xi^\mu\), Eq. \eqref{14} gives \(\delta h_{\alpha\beta}=-\nabla_\alpha\xi_\beta-\nabla_\beta\xi_\alpha\). Substituting this into Eq. \eqref{B.10} and integrating by parts yields
\begin{equation}
\delta_\xi \alpha_A
=\frac{1}{p_0^{\hat{0}}(\lambda_o)}\,
\Pi_{A}{}^{\ \hat{b}}\,e_{\hat{b}}{}^{\ \mu}(x_o)\,
\xi_\mu\Big|_{\lambda_s}^{\lambda_o}
+\mathcal{O}\,\left(\frac{\varepsilon}{r_o}\right), \label{B.15}
\end{equation}
where we used the background geodesic equation and \(\dot p_{0\mu}=0\). With our asymptotic conditions \(\xi^\mu=\mathcal{O}(1/r)\), Eq. \eqref{B.15} reproduces the gauge-independence statement in Eq \eqref{15}: \(\delta_\xi \alpha_A=0+\mathcal{O}(\varepsilon/r_o)\). Thus, the kernel Eq. \eqref{B.11} produces gauge-safe screen observables at first order, provided the observer tetrad is fixed by Eq. \eqref{9}.

In the far-zone reduction implies the \(1/b\) law. Insert the large-\(r\) asymptotics \(A.12\) into \(B.11\), parametrize the background ray by a straight line with longitudinal coordinate \(z\) as in Eq. \eqref{24}, and note that \(p_0^{\hat{0}}(\lambda_o)=E\) to leading order. One finds
\begin{equation}
\delta\alpha_A(b,t_o)=
\mathrm{Re}\left\{e^{-i\omega t_o}\,\frac{c_A(\omega)}{b}\,
\mathcal{F}_A(\omega b)\right\}
+\mathcal{O}\,\left(\frac{M}{b^2}\right), \label{B.16}
\end{equation}
with \(c_A\) and \(\mathcal{F}_A\) defined in Eqs. \eqref{25}-\eqref{26}. This reproduces the weak-field Born expansion used in Sec. \ref{subsec3.2} and provides the link between the kernel normalization and the centroid wobble of Eq. \eqref{30}.

Near-photon-sphere reduction implies the logarithmic law. In the near-critical regime \(b\to b_c^+\), reparametrize the background ray by the azimuth \(\varphi\) accumulated during the whirl [Eq. \eqref{34}]. Use the near-photon-sphere reduction \(A.13\) inside \(B.11\) and integrate over the long dwell time \(\Delta\varphi\sim -\ln\Delta\), with \(\Delta=b/b_c-1\). The integral yields
\begin{equation}
\delta\alpha_A(b,t_o)=
\mathrm{Re}\left\{e^{-i\omega t_o}\,
\left[\mathcal{C}_A(\omega)\,\ln\Delta+\mathcal{D}_A(\omega)\right]\right\}
+\ldots, \label{B.17}
\end{equation}
which is Eq. \eqref{35}. When combined with the background divergence in Eq. \eqref{32} and reorganized as in Eqs. \eqref{36}-\eqref{39}, Eq. \eqref{B.17} fixes the time-dependent SDL coefficients \(a_1,b_1,\beta_1\) from the same kernel.

Starting from Hamilton's equations, we derived the Born response \eqref{B.4} for the photon momentum, projected it to the observer's screen to obtain the master kernel \eqref{B.11}, and exhibited equivalent \(\delta\Gamma\) and delay-space forms, Eqs. \eqref{B.13}-\eqref{B.14}. The kernel is gauge-safe at \(\mathcal{O}(\varepsilon)\) by Eq. \eqref{B.15}. Its far-zone and near-photon-sphere reductions, Eqs. \(B.16\) and \(B.17\), reproduce the weak-field \(1/b\) behavior Sec. \ref{subsec3.2} and the logarithmic SDL structure Sec. \ref{sec4}, respectively. In this sense, all imaging diagnostics discussed in Sec. \ref{sec5} are different asymptotic faces of a single line-of-sight kernel driven by the Zerilli master field.

\section{Near-critical expansion and SDL coefficients} \label{appC}
In this appendix, we provide a self-contained derivation of the near-critical expansion underlying Eqs. \eqref{32} and \eqref{36}-\eqref{39}, and we express the time-dependent strong-deflection-limit \(SDL\) response coefficients \({a_1(\omega),b_1(\omega),\beta_1(\omega)}\) in terms of quantities localized at the photon sphere and of moments of the Born kernel introduced in Appendix \ref{appB}. We keep all conventions of Secs. \ref{sec2}-\ref{sec4}, in particular the Schwarzschild background (Eq. \eqref{1}), the perturbative split (Eq. \eqref{2}), and the master Born representation (Eqs. \eqref{18}-\eqref{22}).

\emph{Static near-critical recap:} For null geodesics on a static, spherically symmetric background \(ds^2=-A(r),dt^2+B(r),dr^2+r^2 d\Omega^2\), the impact parameter as a function of the radial turning point \(r_0\) is
\begin{equation}
b^2(r_0)=\frac{r_0^2}{A(r_0)}. \label{C.1}
\end{equation}
The circular photon orbit \(r_{\mathrm{ph}}\) solves
\begin{equation}
\mathcal{F}(r)\equiv r\,A'(r)-2A(r)=0, \qquad r=r_{\mathrm{ph}}, \label{C.2}
\end{equation}
and the critical impact parameter is \(b_c=r_{\mathrm{ph}}/\sqrt{A(r_{\mathrm{ph}})}\). For Schwarzschild, \(A=f=1-2M/r\) gives \(r_{\mathrm{ph}}=3M\) and \(b_c=3\sqrt{3}M\) as stated in Eq. \eqref{31}. Expanding the background bending integral in Eq. \eqref{16} around \(b\to b_c^+\) yields Eq. \eqref{32},
\begin{equation}
\alpha_0(b)=-\bar a\,\ln\,\left(\frac{b}{b_c}-1\right)+\bar b+\mathcal{O}\,\left((b-b_c)\ln|b-b_c|\right), \label{C.3}
\end{equation}
where \(\bar b\) is a constant depending on the chosen normalization, and $\bar a = 1$ for the Schwarzschild case. Before turning to the time-dependent response, it is useful to state the scope of the derivation precisely. In the present work, the full spacetime is nonstationary because of the QNM perturbation, but the near-critical expansion is performed in a controlled first-order Born treatment: the Schwarzschild background geodesic structure is kept exact, the perturbation is linear in \(\varepsilon\), and the time dependence is evaluated on the unperturbed near-critical ray. Under these assumptions, the standard logarithmic SDL structure survives because it is inherited from the hyperbolic instability of the \emph{background} photon sphere. The role of the perturbation is then to modulate the coefficients \(\bar a,\bar b\) and to shift the critical scale \(b_c\), rather than to replace the logarithm by a different singular structure at \(\mathcal{O}(\varepsilon)\).

\emph{Linear response of the critical scale \(b_c(t_o)\):} We now compute the \(\mathcal{O}(\varepsilon)\) modulation of \(b_c\) induced by the even-parity \((\ell,m)=(2,0)\) perturbation described in Sec. \ref{subsec2.1}. We adopt an instantaneous snapshot characterization at the detection time \(t_o\): the time dependence appears as a harmonic factor \(e^{-i\omega t_o}\) multiplying complex amplitudes evaluated at \(r_{\mathrm{ph}}\). This adiabatic reading is equivalent, to first order, to the near-photon-sphere reduction of the Born kernel and provides a compact route to \(\beta_1(\omega)\).

Let \(A(r)\to A(r)+\varepsilon\,a(r,t)\), where \(A=f\) for the background and \(a\) is the even-parity perturbation of \(g_{tt}\) (we keep \(B\) unmodified since, to first order in \(\varepsilon\), \(b_c\) is controlled by \(A\) alone through Eq. \eqref{C.1}). The photon-sphere condition \eqref{C.2} becomes \(\mathcal{F}(r_{\mathrm{ph}}+\varepsilon\,\delta r)=0\) with
\begin{equation}
\delta r\,\mathcal{F}'(r_{\mathrm{ph}})+\left[r\,a'(r)-2a(r)\right]_{r=r_{\mathrm{ph}}}=0, \notag
\end{equation} 
\begin{equation}
\mathcal{F}'(r_{\mathrm{ph}})=r_{\mathrm{ph}}A''(r_{\mathrm{ph}})-A'(r_{\mathrm{ph}}). \label{C.4}
\end{equation}
Solving Eq. \eqref{C.4} gives
\begin{align}
\delta r &= \frac{2\,a(r_{\mathrm{ph}})-r_{\mathrm{ph}}\,a'(r_{\mathrm{ph}})}{\mathcal{F}'(r_{\mathrm{ph}})} \nonumber \\
&= -\frac{3M}{2}\,\left[2\,a(r_{\mathrm{ph}})-r_{\mathrm{ph}}\,a'(r_{\mathrm{ph}})\right]. \label{C.5}
\end{align}
The fractional change of \(b_c\) follows from \(\ln b_c=\ln r_{\mathrm{ph}}-\frac{1}{2}\ln A(r_{\mathrm{ph}})\):
\begin{equation}
\frac{\delta b_c}{b_c}
=\frac{\delta r}{r_{\mathrm{ph}}}-\frac{1}{2}\left[\frac{a(r_{\mathrm{ph}})}{A(r_{\mathrm{ph}})}+\frac{A'(r_{\mathrm{ph}})}{A(r_{\mathrm{ph}})}\,\delta r\right]. \label{C.6}
\end{equation}
Substituting \(A(r_{\mathrm{ph}})=1/3\), \(A'(r_{\mathrm{ph}})=2/(9M)\), and Eq. \eqref{C.5} leads to a remarkable cancellation of the \(a'(r_{\mathrm{ph}})\) terms and yields
\begin{equation}
\frac{\delta b_c}{b_c}=\frac{3}{2}\,a(r_{\mathrm{ph}}). \label{C.7}
\end{equation}
Thus, the leading near-critical modulation of the critical scale depends only on the instantaneous value of \(a=\delta A\) at the photon sphere. Since \(A=-g_{tt}\), we have \(a=-h_{tt}\). Using the RWZ reconstruction in Eq. \eqref{A.4} with \(m=0\), \(h_{tt}=f,H_0,Y_{20}(\theta)\); evaluating at the equator \(\theta=\pi/2\) and at \(r_{\mathrm{ph}}\) with \(f=1/3\) gives (equator, $P_2$ basis)
\begin{equation}
    \beta_1(\omega)\equiv\frac{\delta b_c}{b_c}=\frac{1}{4}\,H_0(r_{\mathrm{ph}},\omega)
\end{equation} \label{C.8}
In the $Y$-harmonic normalization one has $Y_{20}=\sqrt{\frac{5}{4\pi}}\,P_2$, so $\beta_1=\frac{1}{4}\,H_0^{(P)}=\frac{1}{4}\,\left(\sqrt{\frac{4}{\pi}{5}}\,H_0^{(Y)}\right)$; we use the $P_2$ convention throughout the main text. The time dependence \(\propto \mathrm{Re}{e^{-i\omega t_o}\beta_1(\omega)}\) then reproduces the \(1/\Delta\) term in Eq. \eqref{39}. Eq. \eqref{C.8} is equivalent to extracting the residue of the \(1/\Delta\) pole directly from the Born kernel Appendix \ref{appB}, but makes explicit the quasi-local character of \(\beta_1\).

We comment that the cancellation of \(a'(r_{\mathrm{ph}})\) in Eq. \eqref{C.6} is a special feature of Schwarzschild and the particular combination entering \(b_c\). It reflects the fact that \(b_c\) depends on \(A\) through \(r_{\mathrm{ph}}\) and the normalization \(A(r_{\mathrm{ph}})\) only; at linear order these two contributions conspire to eliminate the derivative at \(r_{\mathrm{ph}}\).

\emph{Logarithmic and finite response: \(a_1(\omega)\) and \(b_1(\omega)\):} 
We next relate the coefficients \(a_1(\omega)\) and \(b_1(\omega)\) in Eq. \eqref{37} to moments of the near-photon-sphere Born kernel. Let \(\hat{\boldsymbol{e}}_b\) be the radial unit vector on the screen. Axisymmetry implies that, to leading order in \(\Delta=b/b_c-1\), the deflection correction \(\delta\boldsymbol{\alpha}\) is radial, so we write \(\delta\boldsymbol{\alpha}=\delta\alpha_{\parallel},\hat{\boldsymbol{e}}_b\). Comparing Eq. \eqref{35} to the scalar linearization \eqref{39} gives
\begin{equation}
a_1(\omega)= -\hat{\boldsymbol{e}}_b\,\cdot\,\boldsymbol{\mathcal{C}}(\omega),
\qquad
b_1(\omega)= \hat{\boldsymbol{e}}_b\,\cdot\,\boldsymbol{\mathcal{D}}(\omega), \label{C.9}
\end{equation}
where \(\boldsymbol{\mathcal{C}}=(\mathcal{C}_X,\mathcal{C}_Y)\) and \(\boldsymbol{\mathcal{D}}=(\mathcal{D}_X,\mathcal{D}_Y)\) are the complex vector amplitudes in Eq. \eqref{35}. In turn, \(\mathcal{C}_A\) and \(\mathcal{D}_A\) follow from the near-photon-sphere reduction of the kernel \(B.11\). Introducing the azimuth \(\varphi\) during the whirl phase (Eq. \eqref{34}) and the delay \(\tau(\varphi)\), we can write
\begin{equation}
\mathcal{C}_A(\omega)=\int^{\varphi_{\max}}_{\varphi_{\min}} d\varphi\,W_A^{(0)}(\varphi),\notag 
\end{equation} 
\begin{equation}
\mathcal{D}_A(\omega)=\int^{\varphi_{\max}}_{\varphi_{\min}} d\varphi\,\left[W_A^{(1)}(\varphi)
+ i\omega\,U_A^{(1)}(\varphi)\right], \label{C.10}
\end{equation}
where the weights \(W_A^{(0)}\), \(W_A^{(1)}\), \(U_A^{(1)}\) are smooth functions of the background quantities at \(r\simeq r_{\mathrm{ph}}\) and of the RWZ amplitudes \(\Psi_Z\,\partial_{r_*}\Psi_Z\) evaluated there (see Eqs. \eqref{A.13} and \eqref{B.17}). Their explicit algebraic forms are lengthy but completely fixed by Eqs. \eqref{A.4}, \eqref{A.8}-\eqref{A.11}, and \eqref{B.11}. The limits \((\varphi_{\min},\varphi_{\max})\) scale as \(|\ln\Delta|\) and generate the logarithm in Eq. \eqref{35}; the finite parts define \(b_1\) once the inner and outer contributions are combined as in Eq. \eqref{50}.

Eqs. \eqref{C.9}-\eqref{C.10} summarize the content of \(a_1\) and \(b_1\): both are localized, frequency-resolved moments of the Zerilli field and its first radial derivative at \(r_{\mathrm{ph}}\), with \(a_1\) capturing the universal logarithmic weight and \(b_1\) the finite near-photon-sphere remainder after matching to the outer integral (Sec. \ref{subsec5.1}).

\emph{Time-of-flight counterparts:} The coordinate travel time for near-critical rays has the structure as Eq. \eqref{43},
\begin{align}
T(b,t_o)&=-\tilde a(t_o)\,\ln\,\left(\frac{b} {b_c(t_o)}-1\right) \nonumber \\ 
&+\tilde b(t_o) 
+n\,T_{\mathrm{whirl}}(t_o)+\ldots, \label{C.11}
\end{align}
where \(n\) counts azimuthal winds. Repeating the steps that led to \(C.9\) with the temporal kernel (obtained by replacing the screen projector in Eq. \eqref{B.11}) with the unit covector \(u_\mu=e^{\hat{0}}{}_\mu)\) gives
\begin{equation}
\tilde a_1(\omega)= -\hat{\boldsymbol{t}}\,\cdot\,\boldsymbol{\mathcal{C}}^{(T)}(\omega),\qquad
\tilde b_1(\omega)= \hat{\boldsymbol{t}}\,\cdot\,\boldsymbol{\mathcal{D}}^{(T)}(\omega),\notag 
\end{equation} 
\begin{equation}
\gamma_1(\omega)=\frac{\delta T_{\mathrm{whirl}}}{T_{\mathrm{whirl}}}, \label{C.12}
\end{equation}
where \(\hat{\boldsymbol{t}}\) denotes projection onto the time leg of the tetrad and the superscript \(T\) reminds us that the weights are those appropriate to the time functional. The same near-photon-sphere data \({\Psi_Z\,\partial_{r_*}\Psi_Z}_{r_{\mathrm{ph}}}\) enter, so the inter-image delay modulation (Eq. \eqref{46}) and the spacing/ring modulations (Eqs. \eqref{47}-\eqref{48}) are tied to the same localized kernel, as emphasized in Sec. \ref{subsec5.2}.

\emph{Consistency with the \(1/\Delta\) pole and matching:} Expanding the time-dependent law \(36\) to linear order using \(b_c(t_o)=b_c,[1+\varepsilon\,\mathrm{Re}{e^{-i\omega t_o}\beta_1(\omega)}]\) and \(\bar a(t_o)=\bar a+\varepsilon\,\mathrm{Re}{e^{-i\omega t_o}a_1(\omega)}\) reproduces Eq. \eqref{39} with the identifications
\begin{align}
    \text{(i)}\  &\beta_1(\omega)=\frac{\delta b_c}{b_c}\quad \text{from (C.8)},\nonumber \\
\text{(ii)}\  &a_1(\omega),b_1(\omega)\ \text{ from (C.9)-(C.10)}.
\end{align}
The \(1/\Delta\) term \(\bar a,\beta_1/\Delta\) is fixed entirely by \eqref{C.8} and is independent of the arbitrary split radius used in Eq. \eqref{49}; the logarithmic and finite pieces are determined by the inner kernel moments and are rendered unique by the matching cancellations between Eqs. \eqref{50}-\eqref{51}. This is the precise content of the overlap relation \eqref{53}: the near-critical asymptotics of the \(1/b\) Born amplitude encodes the triplet \((\beta_1,a_1,b_1)\).

Near criticality, all time-dependent SDL coefficients are controlled by local data at the photon sphere together with universal whirl-integrals of the Born kernel. In particular: (i) the critical-scale modulation is quasi-local,
\begin{equation}
\beta_1(\omega)=\frac{1}{4}\,H_0(r_{\mathrm{ph}},\omega),
\end{equation}
equivalently \(\delta b_c/b_c=(3/2)\,a(r_{\mathrm{ph}})\) with \(a=-h_{tt}\) and \(P_2(0)=-1/2\); (ii) the logarithmic and finite coefficients \(a_1(\omega)\) and \(b_1(\omega)\) are given by kernel moments in Eq. \eqref{C.10} built from \(\Psi_Z\) and \(\partial_{r_*}\Psi_Z\) at \(r_{\mathrm{ph}}\), projected radially according to Eq. \eqref{C.9}; and (iii) the time-of-flight counterparts \(\tilde a_1,\tilde b_1,\gamma_1\) arise from the same local ingredients via Eq. \eqref{C.12}, explaining the phase-locked modulations of spacing and delays (Sec. \ref{subsec4.2}) and their correlation with ring-radius oscillations (Sec. \ref{subsec5.2}).

This also clarifies the central claim of the manuscript: the weak-field \(1/b\) behavior and the near-critical logarithmic SDL structure are not derived from separate constructions, but from different asymptotic reductions of the same line-of-sight kernel. The far-zone reduction gives the weak-field amplitude, while the near-photon-sphere reduction yields the logarithmic SDL coefficients and the critical-scale shift. The matching relation of Sec. \ref{subsec5.1} is therefore the precise statement that the near-critical triplet \((\beta_1,a_1,b_1)\) is encoded in the same Born kernel that governs the weak-field response.

These results complete the derivation of Eqs. \eqref{36}-\eqref{39} from the kernel formalism, making explicit why the logarithmic structure persists at first order in the time-dependent setting considered here and how the Zerilli master field, evaluated at the photon sphere, controls all near-critical, time-dependent imaging phenomenology in our unified framework.

\acknowledgments
The author is grateful for helpful discussions with A. {\"O}vg{\"u}n. R. P. would like to acknowledge networking support of the COST Action CA21106 - COSMIC WISPers in the Dark Universe: Theory, astrophysics and experiments (CosmicWISPers), the COST Action CA22113 - Fundamental challenges in theoretical physics (THEORY-CHALLENGES), the COST Action CA21136 - Addressing observational tensions in cosmology with systematics and fundamental physics (CosmoVerse), the COST Action CA23130 - Bridging high and low energies in search of quantum gravity (BridgeQG), the COST Action CA23115 - Relativistic Quantum Information (RQI) funded by COST (European Cooperation in Science and Technology), and SCOAP3 (Switzerland) for their support.

\bibliography{ref}

\end{document}